\documentclass[]{aastex}

\newcommand{\myemail}{cwalsh@strw.leidenuniv.nl}

\usepackage{amsmath}
\usepackage{amssymb}
\usepackage{url}
\usepackage{gensymb}
\usepackage{graphics}
\usepackage{subfigure}
\usepackage{wasysym}
\usepackage{color}

\shorttitle{Molecular Line Emission from an Externally Irradiated Protoplanetary Disk}
\shortauthors{Walsh, Millar, \& Nomura}

\begin{document}

\title{MOLECULAR LINE EMISSION FROM A PROTOPLANETARY DISK IRRADIATED EXTERNALLY BY A NEARBY MASSIVE STAR}

\author{Catherine Walsh\altaffilmark{1,2}, T. J. Millar\altaffilmark{1}, \& Hideko Nomura\altaffilmark{3}} 
\altaffiltext{1}{Astrophysics Research Centre, School of Mathematics and Physics, Queen's University
Belfast, University Road, Belfast BT7 1NN, UK}
\altaffiltext{2}{Leiden Observatory, Leiden University, P.O. Box 9513, NL-2300 RA Leiden, The Netherlands}
\altaffiltext{3}{Department of Astronomy, Graduate School of Science, Kyoto University, Kyoto
606-8502, Japan}
\email{\myemail}
\received{2012 December 28}
\accepted{2013 March 4}

\begin{abstract}
Star formation often occurs within or nearby stellar clusters.  
Irradiation by nearby massive stars can photoevaporate protoplanetary disks around 
young stars (so-called proplyds) which raises questions regarding the ability of planet formation 
to take place in these environments.  
We investigate the two-dimensional physical and chemical structure of a protoplanetary disk 
surrounding a low-mass (T~Tauri) star which is irradiated by a nearby massive O-type star 
to determine the survivability and observability of molecules in proplyds.  
Compared with an isolated star-disk system, the gas temperature ranges from a factor of a few 
(in the disk midplane) to around two orders of magnitude (in the disk surface) higher in the 
irradiated disk.  
Although the UV flux in the outer disk, in particular, is several orders of magnitude higher, 
the surface density of the disk is sufficient for effective shielding of the disk midplane so that 
the disk remains predominantly molecular in nature.  
We also find that non-volatile molecules, such as HCN and H$_2$O, 
are able to freeze out onto dust grains in the disk midplane so that 
the formation of icy planetesimals, e.g., comets, may also be possible in proplyds.  
We have calculated the molecular line emission from the disk assuming LTE 
and determined that multiple transitions of atomic carbon, 
CO (and isotopologues, $^{13}$CO and C$^{18}$O), 
HCO$^+$, CN, and HCN may be observable with ALMA, allowing characterization of 
the gas column density, temperature, and optical depth in proplyds at 
the distance of Orion ($\approx$400~pc).  
\end{abstract}


\section{INTRODUCTION}
\label{introduction}

Star formation rarely takes place in isolation;   
stars form primarily within multiple systems and  
often within, or nearby, massive stellar clusters 
\citep[see, e.g.,][]{kennicutt12}.  
Thus, the young star's 
protoplanetary disk can be heavily irradiated and indeed, photoevaporated, by 
nearby massive stars \citep[see, e.g.,][]{bally98}, for example,   
the famous {\em proplyds} in the Orion nebula \citep[][]{odell93}.  
The properties of proplyds have been well studied since their discovery;  
however, little is known 
about their molecular content due to their small size 
($R$~$\sim$~100--1000~AU) and relatively large distance, 
e.g., Orion at $\approx$400~pc.  
Protoplanetary disks are the sites of planet formation and 
the study of disks in extreme environments 
is important for a complete understanding of this process.  
Can the disk survive sufficiently long for planet formation to take place?
This has implications on our understanding of the formation of our own
solar system as it is thought that the Sun formed within a 
stellar cluster which has long since dispersed 
\citep[see, e.g.,][and references therein]{adams10}.

We calculate the chemical structure and molecular line emission 
from a protoplanetary disk around a classical T~Tauri star 
which is irradiated by a nearby ($\lesssim$0.1~pc) massive O-type star.    
Our aims are (1) to determine whether molecules are able to form and survive 
in the disk, (2) to identify species which may be observable with ALMA, and  
(3) to identify species which can characterize the physical conditions and the subsequent 
implications on planet formation.

\section{PROTOPLANETARY DISK MODEL}
\label{diskmodel}

\subsection{Physical Model}
\label{physicalmodel}

We model a steady axisymmetric disk in Keplerian rotation about a classical 
T~Tauri star with mass 0.5~$M_\odot$, radius 2~$R_\odot$, effective 
temperature 4000~K, and $\alpha$~=~0.01, using the methods outlined in 
\citet{nomura05} and \citet{nomura07}. 
The radial surface density distribution is determined 
assuming a constant mass accretion rate, 
$\dot{M}$~=~10$^{-8}$~$M_\odot$~yr$^{-1}$ \citep{pringle81}.
To determine the density and temperature structure, 
we solve the equation of hydrostatic equilibrium in the 
vertical direction and the local 
thermal balance between the heating and cooling of the gas. 
We include grain photoelectric heating by UV photons and heating 
via hydrogen ionization by X-rays and  
allow the gas to cool via gas--grain collisions and line transitions.  

Our disk is irradiated by UV photons and X-rays from the central star 
and UV photons from the interstellar radiation field and a nearby massive 
O-type star.  
We model the UV spectrum of the O-type star as a blackbody 
with an effective temperature of 45,000~K.
We set the UV flux at the disk surface to 
(4~$\times$~10$^{5}$)~$\times$~$G_{0}$, where 
$G_0$~$\approx$~1.6~$\times$~10$^{-3}$~erg~cm$^{-2}$~s$^{-1}$ 
is the average strength of the integrated interstellar radiation field.  
This corresponds to a distance between the disk and the massive star of $\lesssim$0.1~pc.  

We assume the dust and the gas in the disk are well mixed and adopt 
a dust-grain size distribution which reproduces the extinction curve observed 
in dense clouds \citep{weingartner01}.  
We note that this is a simplistic treatment of the dust distribution in 
disks since it is thought that gravitational dust settling and coagulation perturb 
the dust size and density distribution from that observed in dense clouds 
\citep[see, e.g.,][]{dullemond04,dalessio06}.  

\subsection{Chemical Model}
\label{chemicalmodel}

We use the gas--grain chemical network from 
\citet{walsh10} and compute the photoreaction rates and X-ray ionization rates
using the methods outlined in \citet{walsh12}. 
Our gas-phase chemistry is from the UMIST Database for Astrochemistry
\citep{woodall07}.  
We include freezeout of gas-phase material onto 
dust grains and thermal and non-thermal desorption.  
We allow photodesorption by both external and internal 
UV photons 
and desorption via cosmic-ray heating of dust grains 
\citep{hasegawa93,oberg09a,oberg09b}.  
We use a small grain-surface network to include the synthesis of several 
complex organic molecules, e.g., methanol \citep{hasegawa92,hasegawa93}.  
We include the photodestruction of grain mantle material by 
UV photons, X-rays, and cosmic rays \citep{ruffle01}.  
Experiments have shown that the 
irradiation of interstellar ice analogs
by UV and X-ray photons induces chemistry in the ice 
\citep[see, e.g.,][]{hagen79,dhendecourt86,allamandola88,oberg10a,ciaravella10}. 
We adopt the equivalent gas-phase photorates 
for ice species and allow the products to remain on the 
grain to take part in further grain-surface reactions.  

We assume the grains are negatively charged 
compact spheres with a radius of 0.1~$\mu$m 
and a constant fractional abundance of $\sim$10$^{-12}$ relative to the 
gas number density, equivalent to a gas-to-dust mass ratio of $\sim$100.    
We adopt a sticking coefficient, $S$~$\approx$~1, for all species.  
We include the dissociative recombination of gas-phase cations with dust grains. 
The dust model used for the chemistry is simple compared with 
that used to calculate the disk physical structure; 
however, the total dust-grain cross section is consistent between models. 

Our initial abundances are from the output of a dark cloud model 
at an age of $\sim$10$^5$~yr.
Our underlying elemental abundance ratios for 
H:He:C:N:O:Na:Mg:Si:S:Cl:Fe are
1.00:0.14:7.30(--5):2.14(--5):1.76(--4):3.00(--9):3.00(--9):3.00(--9):2.00(--8):3.00(--9):3.00(--9).  
We map the chemical structure of the disk at a time of $10^6$~yr, 
the typical age of classical T~Tauri stars.  

\subsection{Molecular Line Emission}
\label{radiativetransfer}

To estimate the molecular line emission, we assume the disk is ``face on" 
and that local thermodynamic equilibrium (LTE) holds throughout.
We calculate the emergent flux from the top (i.e., visible) half of the 
disk only which we also assume is the surface directly irradiated by the 
nearby massive star. 
In reality, the lower half, although not directly irradiated by the nearby star,
is subject to diffuse, scattered UV radiation 
originating from the stellar cluster, and this needs to be taken into account 
in the calculation of the physical structure of the lower disk.  
Hence, our estimated molecular line emission strengths are 
{\em lower limits} to the potentially observable line strengths.  
We include only thermal broadening of the line profiles because 
this will dominate turbulent broadening.   
For CO isotopologues, we assume the molecular isotopic ratios reflect the 
atomic ratios and use values applicable for the local interstellar medium: 
$^{13}$C/$^{12}$C~=~1/77 and $^{18}$O/$^{16}$O~=~1/560 \citep{wilson94}.  

\section{RESULTS}
\label{results}

\subsection{Physical Structure}
\label{physicalstructure}

In Figure~\ref{figure1}, we present the gas temperature,  
number density, integrated UV flux, 
and integrated X-ray flux as a function of disk radius, $R$, and height, $Z/R$.  
We also display these parameters, along with the cosmic-ray and X-ray ionization 
rates, as a function of height at $R$~=~100~AU. 
We compare results from our irradiated disk (solid lines) and isolated disk 
\citep[dashed lines; see][]{walsh12}.  
Here, ``irradiated" refers to a disk irradiated by a nearby massive star 
{\em and} the central star, whereas, ``isolated" refers to a disk irradiated 
by the central star {\em only}.

Compared with the isolated disk, the gas and dust temperatures in the 
irradiated disk are significantly higher in the outer disk midplane 
($R$~$\gtrsim$~1~AU).  
The gas temperature structure in the innermost region, 
$R$~$\lesssim$~1~AU, is similar in both disks because this is 
driven by irradiation by the central star.  
The density structure is also perturbed, depending on the 
gas temperature profile.   

In the surface at 100~AU, we see around an order of magnitude decrease in the 
X-ray flux and subsequent ionization rate in the irradiated model, 
compared with the isolated model.  
This is due to the differing attenuation of X-rays along the line of sight from the 
central star, which is affected by the density and gas temperature profile in the disk.  
In the disk midplane, the X-ray flux (and ionization rate) is similar in both models.    
We see a large increase in the UV flux in the disk surface, increasing from a value of 
$\sim$1~erg~cm$^{-2}$~s$^{-1}$ in the isolated case to 
$\sim$100~erg~cm$^{-2}$~s$^{-1}$ in the irradiated case at $R$~=~100~AU. 
The surface density in the irradiated disk is sufficient for the
disk midplane to remain effectively shielded from all sources of UV 
irradiation.

\subsection{Chemical Structure}
\label{chemicalstructure}

In Figure~\ref{figure2}, we present the fractional abundance 
(relative to number density) of gas-phase
CO, HCO$^+$, CN, HCN, C$_2$H, CS, H$_2$CO and N$_2$H$^+$ as 
a function of disk radius, $R$, and height, $R$/$Z$.  
These species have been detected in numerous 
nearby gas-rich primordial isolated protoplanetary disks 
due their relatively large abundance and simple rotational spectra 
leading to strong emission lines 
\citep[see, e.g.,][]{thi04,dutrey07,henning10,oberg10b,dutrey11}.  
In Figure~\ref{figure3}, we present the vertical column densities 
as a function of disk radius, $R$. 
We have smoothed the column densities using a B\'{e}zier function 
of index, $n$, where $n$ is the number of data points.  
We also present the column densities for those additional species 
observed in protoplanetary disks in absorption and/or emission at infrared 
wavelengths: OH, H$_2$O, C$_2$H$_2$, and CO$_2$ 
\citep[bottom, see, e.g.,][]{lahuis06,carr08,mandell12}.  
We also show the grain-surface column densities (dashed lines) for species 
which exhibit significant freezeout.  
In Table~\ref{table1}, we list the calculated column densities at radii of 
10~AU and 100~AU.  

We see a different chemical stratification in the irradiated 
disk, compared with the isolated disk.  
The higher dust temperature in the disk midplane allows 
volatile molecules, such as CO, 
to remain in the gas, whereas tightly bound molecules, such as HCN, 
freeze onto grain surfaces.  
Instead of the usual ``warm molecular layer" observed 
between 0.3~$\lesssim$~$Z/R$~$\lesssim$~0.5,
we see a thinner molecular layer which is deeper in the 
disk and abundant in radicals, e.g., CN, C$_2$H, and CS, and  
molecules.  
HCO$^+$ resides in a layer slightly higher than the ``molecule/radical" layer
reflecting the abundance of proton-donating molecules, such as, 
H$_3$$^+$ and H$_3$O$^+$. 

Gas-phase CO possesses its canonical fractional 
abundance of 10$^{-4}$ throughout the 
disk midplane and also survives in the disk surface beyond 
$R$~$\gtrsim$~10~AU with a fractional abundance, $\sim$10$^{-7}$.  
Thus, the disk remains significantly molecular in nature, even under extreme 
irradiation.    
We see hints of the truncation of the molecular disk 
beyond $\approx$50~AU. 
Here, we have assumed hydrostatic equilibrium, whereas, 
in reality, photoevaporative flow may affect the surface density of the 
outer disk. 
The disk truncation radius depends primarily on the scale of the flow 
\citep[see, e.g.,][]{adams04}.

HCN is depleted in the disk midplane due to freezeout and 
reaches a fractional abundance of $\sim$10$^{-7}$ in the 
``molecule/radical" layer beyond $R$~=~1~AU.  
Within this radius, thermal desorption from dust grains 
increases the fractional abundance to $\sim$10$^{-4}$. 
HCO$^+$ and CN both reach a fractional abundance of $\sim$10$^{-6}$, whereas 
C$_2$H and CS reach a value of $\sim$10$^{-7}$.  
H$_2$CO and N$_2$H$^+$ reach lower peak values of 
$\sim$10$^{-8}$ and $\sim$10$^{-9}$, respectively.  
N$_2$H$^+$ reaches this value within a radius $\lesssim$1~AU only. 

Comparing column densities at 10~AU, 
there are minor differences between models 
(less than a factor of three) for most species. 
However, in the irradiated disk, 
the CN, CS, and CO$_2$ column densities are larger, 
by factors of 4.7, 44, and 1800, 
respectively, and those 
for N$_2$H$^+$, H$_2$O, and C$_2$H$_2$, 
are smaller, by factors of 
9.6, 7.0, and 9.6, respectively. 
The CN/HCN, OH/H$_2$O, C$_2$H/C$_2$H$_2$ column density ratios are also larger in the 
irradiated disk, 1.9, 2.9, and 8.2 versus 0.16, 0.26, and 0.29, respectively, 
indicative of increased photodissociation of HCN, H$_2$O, and C$_2$H$_2$.
Both models exhibit peaks in the CS column density 
where the dust temperatures are comparable 
(10~AU in the irradiated disk and 4-5~AU in the isolated disk.)
Hence, freezeout onto dust grains accounts for the difference in the 
CS column densities at 10~AU.  
The CO$_2$ column density is significantly larger in the irradiated disk
$\approx$2~$\times$~10$^{19}$~cm$^{-2}$ versus $\sim$10$^{16}$~cm$^{-2}$. 
In the isolated disk, the snow line for CO$_2$ lies within 
10~AU; however, in the irradiated disk, the temperature in the midplane 
at 10~AU, and beyond, is high enough for CO$_2$ to possess a significant 
gas-phase abundance.

At 100~AU, the effects of external irradiation are clear.   
The H/H$_2$ column density ratio is much larger in the irradiated disk, 
2.9~$\times$~10$^{-2}$ versus 
6.5~$\times$~10$^{-5}$ for the isolated disk.  
The OH/H$_2$O and C$_2$H/C$_2$H$_2$ column density 
ratios are also larger, 3.3 and 7.0 versus 
0.20 and 0.47, respectively. 
The OH and H$_2$O column densities also  
increase significantly in the irradiated disk, 
by factors of 94 and 5.8, respectively. 
OH and H$_2$O can form efficiently in warm/hot gas  
($T$~$\gtrsim$~200~K) via the reactions, 
H$_2$~+~O and H$_2$~+~OH \citep{glassgold09}.
The HCO$^+$ column density is a factor of 6.3 larger in the irradiated disk, 
tracing the increased ionization degree.  
We see only minor differences 
in the column densities of all other species, except CO$_2$ (already discussed) 
and N$_2$H$^+$, the latter of which is a factor of 35 smaller 
in the irradiated disk, at 100~AU. 
N$_2$H$^+$ is destroyed via reactions with CO and electrons.  
The CO abundance in both models is similar; however, 
the electron abundance in the molecular layer of the irradiated model 
is several orders of magnitude higher so that N$_2$H$^+$ destruction via 
dissociative recombination is more effective in this model.  

The best astrophysical analog of our isolated disk model is TW~Hya.  
We achieve good agreement with the disk-averaged abundance ratios for 
CN/HCN derived by \citet[][7.1 versus 3.8 in our model]{thi04}  
and the fractional abundance of water vapor 
determined by 
\citet[][5-20~$\times$~10$^{-8}$ versus 1-3~$\times$~10$^{-8}$]{hogerheijde11}. 
Adjusting for source size and distance, 
we achieve reasonable agreement with H$_2$CO line strengths 
observed by \citet{qi13}: 
1.22 and 0.54~Jy~km~s$^{-1}$ versus 0.65 and 0.82~Jy~km~s$^{-1}$ from our model 
for the 4$_{14}$--3$_{13}$ and 5$_{15}$--4$_{14}$ transitions, 
respectively (C. Walsh et al., in preparation).
We get poor agreement for the HCO$^+$/CO ratio from 
\citet[][1.4~$\times$~10$^{-4}$ versus 2.5~$\times$~10$^{-6}$]{thi04} 
and our N$_2$H$^+$ line strengths are around an order of magnitude 
lower than that observed. 
The truncation of our disk at 100~AU means we do not account 
for the depletion of CO onto dust grains at larger radii 
where $T_\mathrm{dust}$~$\lesssim$~20~K. 
As discussed previously, N$_2$H$^+$ is destroyed via reaction with CO.

\subsection{Molecular Line Emission}
\label{molecularlineemission}

In Figure~\ref{figure4}, we present our synthetic line spectra in ALMA 
receiver bands 6--9.  
We assume a maximum disk radius of 
100~AU and a distance to source of 400~pc.  
The lines in bands 3 and 4 
are likely too weak to be observable 
and observations in band 10 are difficult due to 
increased atmospheric absorption at higher frequencies.  
Due primarily to the higher gas temperature, 
the emission lines from the irradiated 
disk reach higher {\em peak} values than the equivalent lines from the isolated disk.  
This is most notable for C, CO (and isotopologues), CN, CS, and HCN.  
The HCO$^+$ transitions in bands 6 and 7 reach higher 
peak flux densities in the isolated disk 
and are also much stronger than those for HCN, 
despite both molecules possessing similar rotational 
spectra.  
HCO$^+$ is a tracer of cold, dense material and is generally 
more abundant than HCN in the outer disk midplane.

In Table~\ref{table1}, we list the transitions, frequencies, and 
integrated line intensities. 
Due to the higher gas temperature in the irradiated disk ($\gg$50~K),
lower frequency transitions ($\nu$~$\sim$~100~GHz) are weaker than 
higher frequency transitions and the integrated line strengths are 
stronger than those for the isolated disk.  
A significant proportion of the CO,$^{13}$CO, and C$^{18}$O ladders 
should be observable with ALMA advantageously allowing characterization 
of the gas excitation temperature, column density, and optical depth  
\citep[see, e.g.,][]{bruderer12}.  
Multiple transitions of CI, HCO$^+$, HCN, and CN may also be observable 
allowing additional characterization of the physical conditions.  
The CN/HCN and CI/CO line ratios and derived column densities 
would also allow a measurement of the degree of 
photodestruction in the disk.  
The remaining molecules would be difficult to observe; 
however, CS and C$_2$H may be observable in larger, more massive, proplyds 
(100~AU~$\ll$~$R$~$\lesssim$~400~AU).

\section{DISCUSSION}
\label{discussion}

\begin{figure*}
\includegraphics[width=0.5\textwidth]{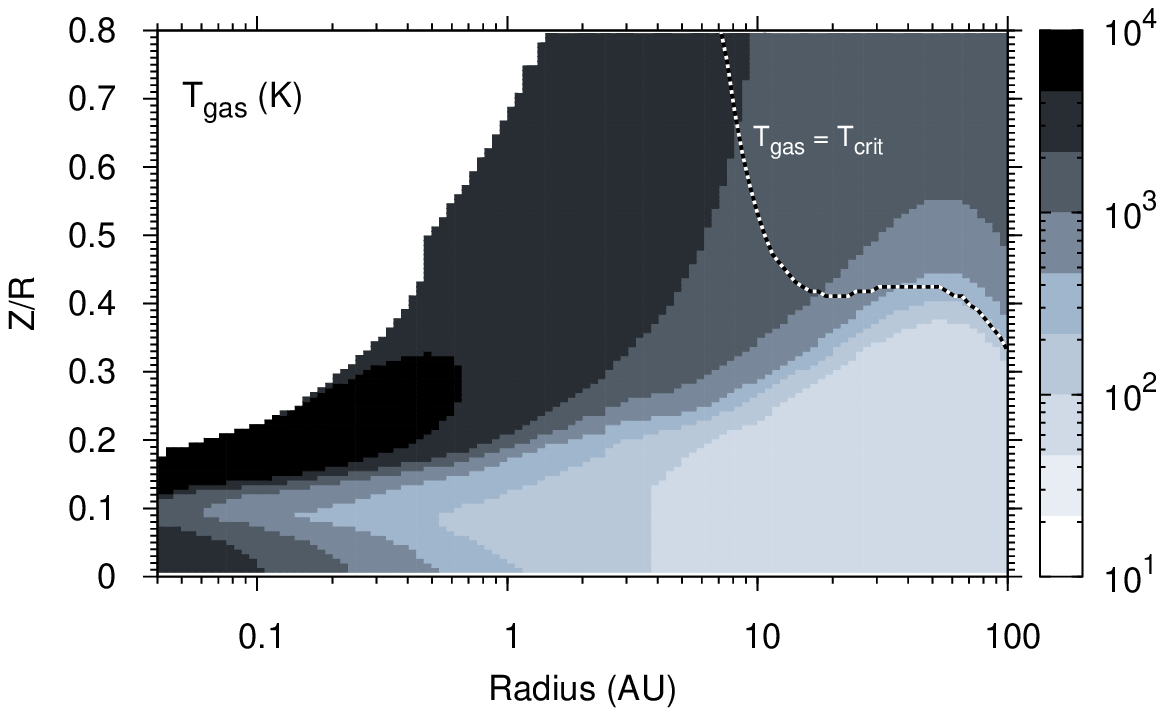}
\includegraphics[width=0.5\textwidth]{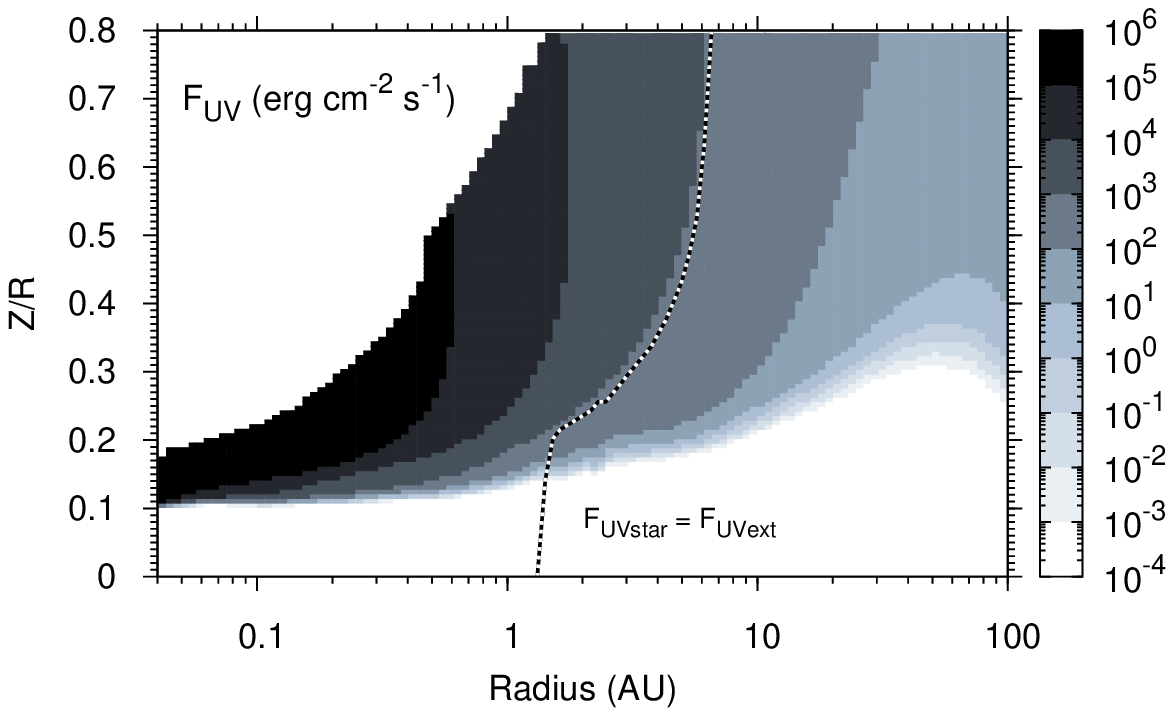}
\includegraphics[width=0.5\textwidth]{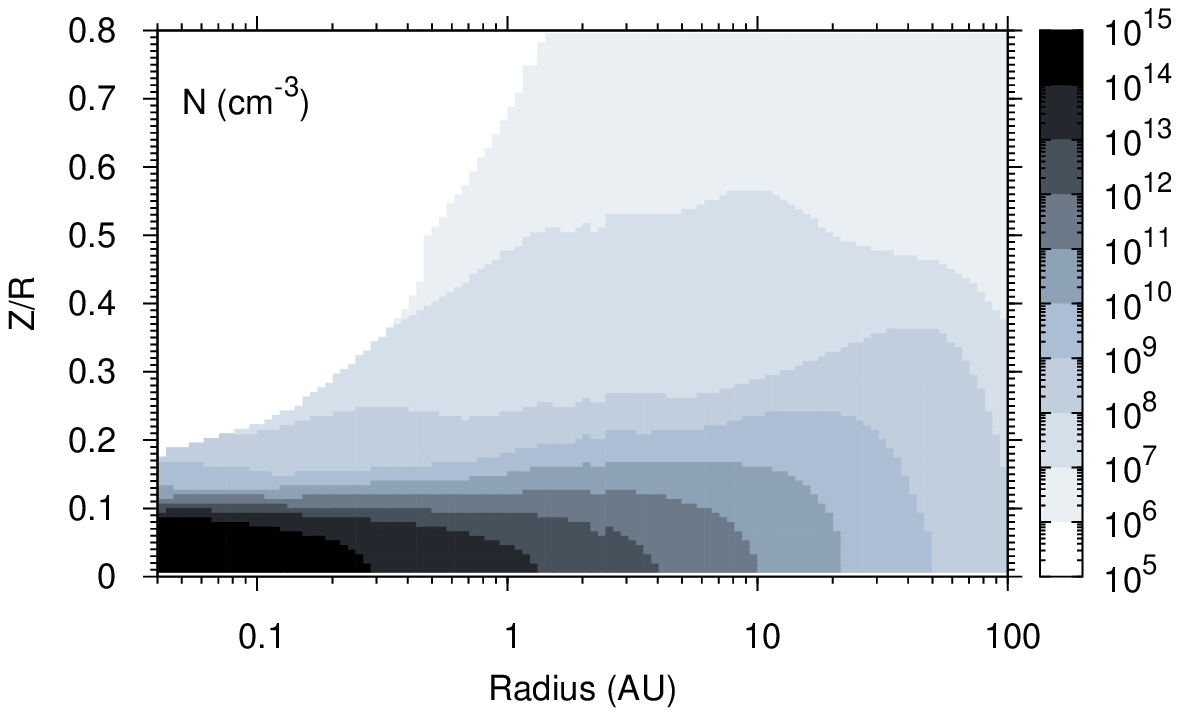}
\includegraphics[width=0.5\textwidth]{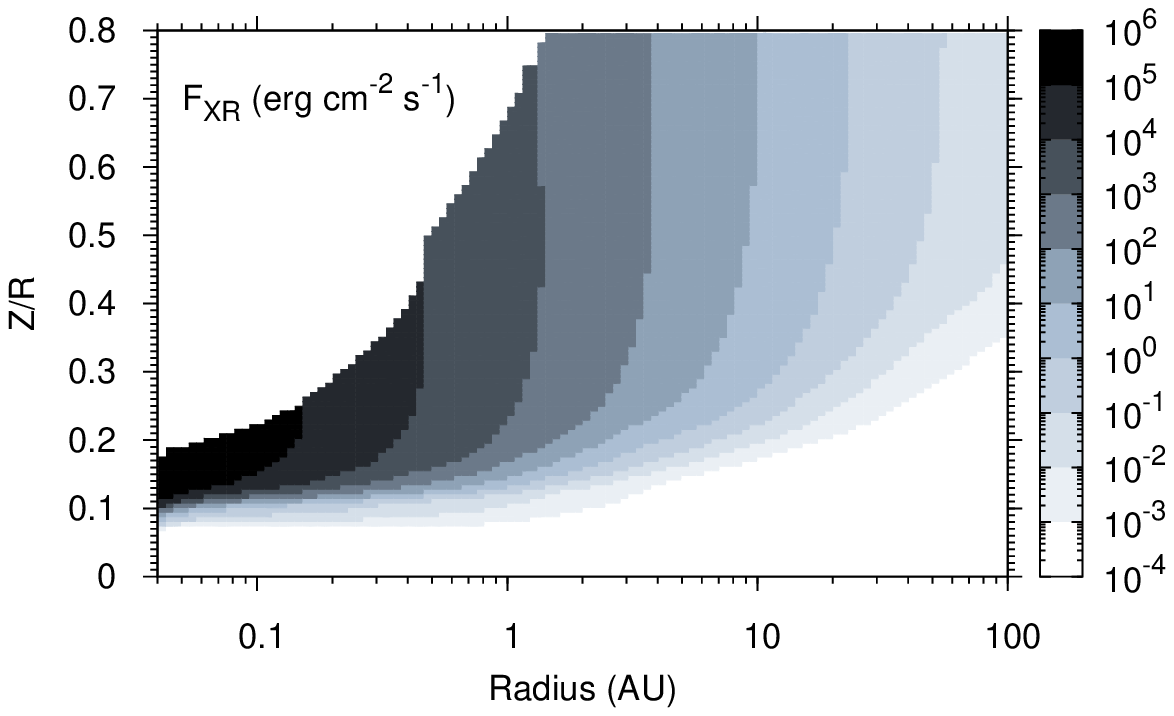}
\includegraphics[width=0.5\textwidth]{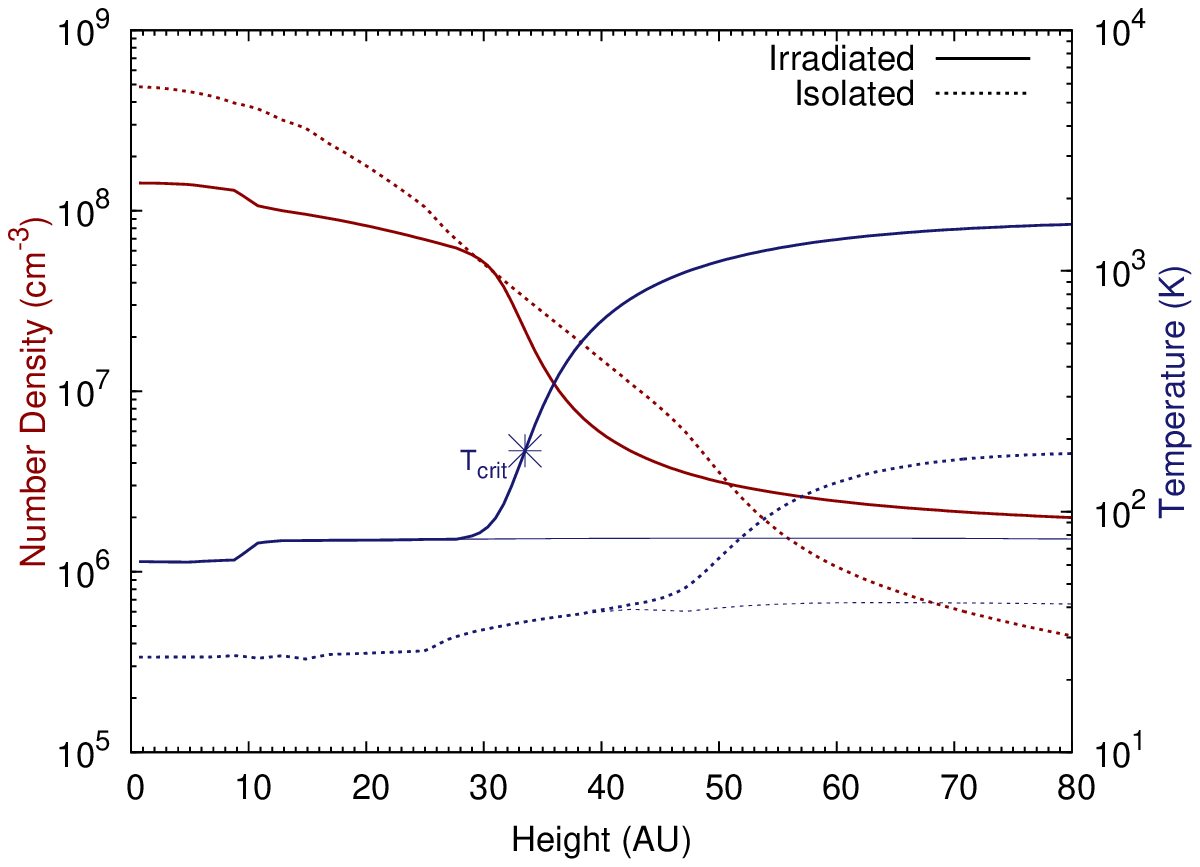}
\includegraphics[width=0.5\textwidth]{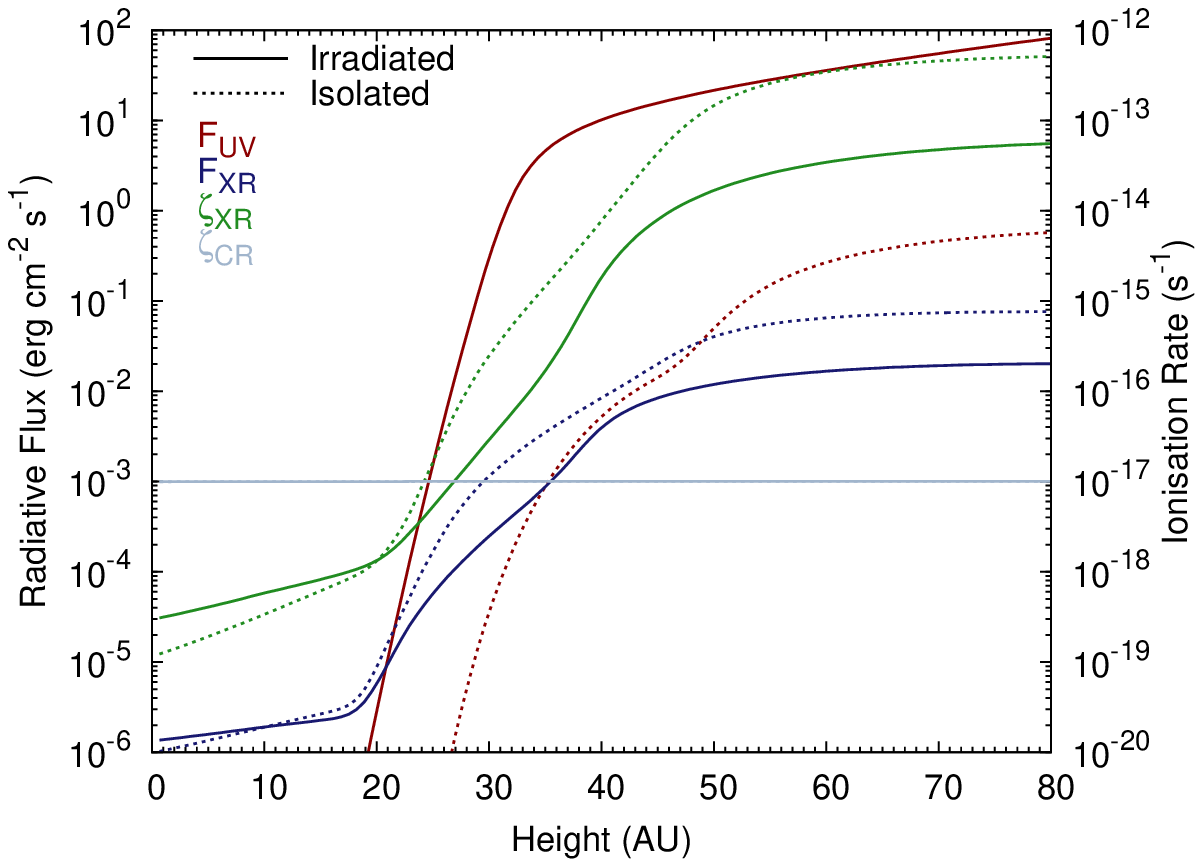}
\caption{Gas temperature, number density, 
UV flux and X-ray flux as a function of disk 
radius, $R$, and height, $Z/R$ (top two rows).  
Gas and dust temperature, number density, UV flux, X-ray flux, 
X-ray ionization rate, and cosmic-ray ionization rate as a function of height, $Z$, 
at a radius of 100~AU (bottom row), for an irradiated disk (solid lines) and an 
isolated disk (dashed lines).  
The lines in the top left and right panels identify where the 
gas temperature equals the critical temperature for photoevaporation, $T_\mathrm{crit}$, and 
where the contribution to the UV field strength from the central star equals that from 
the external star.}
\label{figure1}
\end{figure*}

\begin{figure*}
\includegraphics[width=0.5\textwidth]{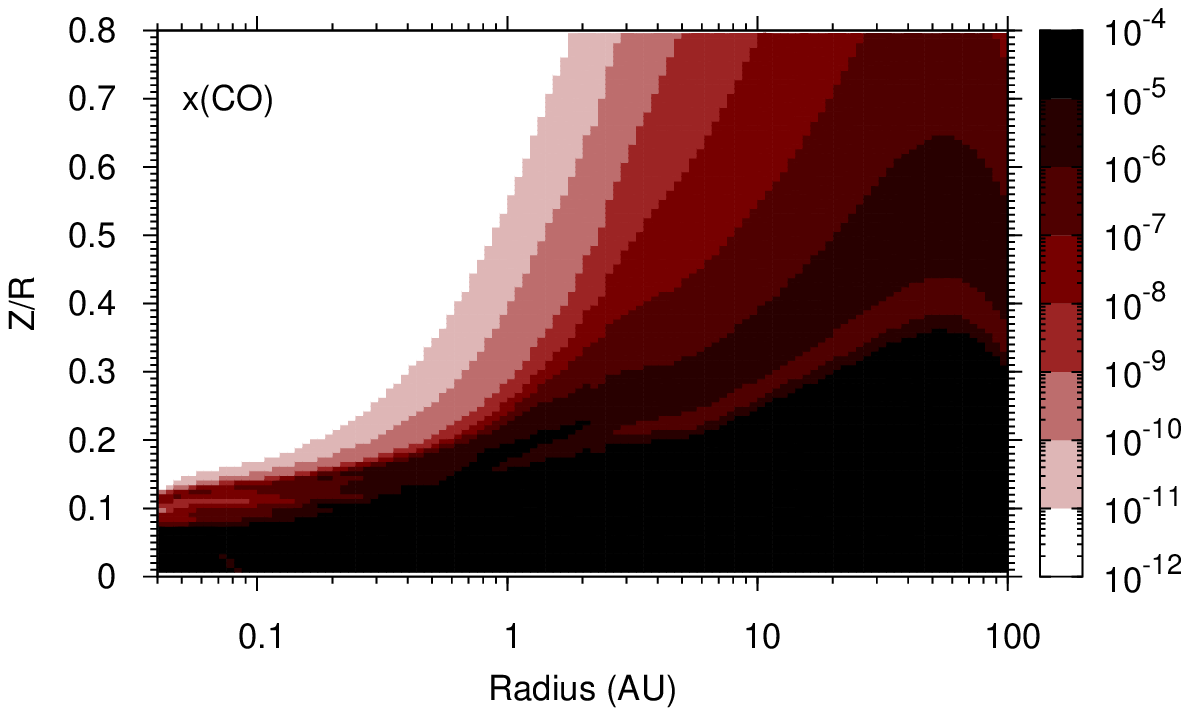}
\includegraphics[width=0.5\textwidth]{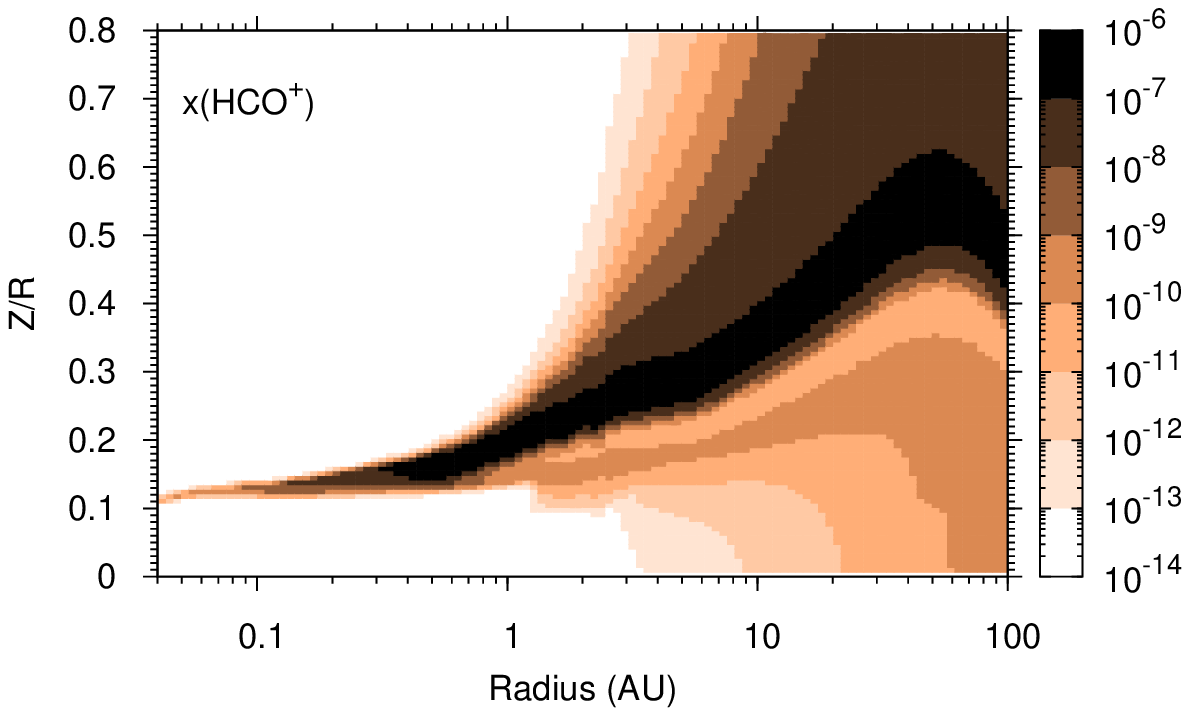}
\includegraphics[width=0.5\textwidth]{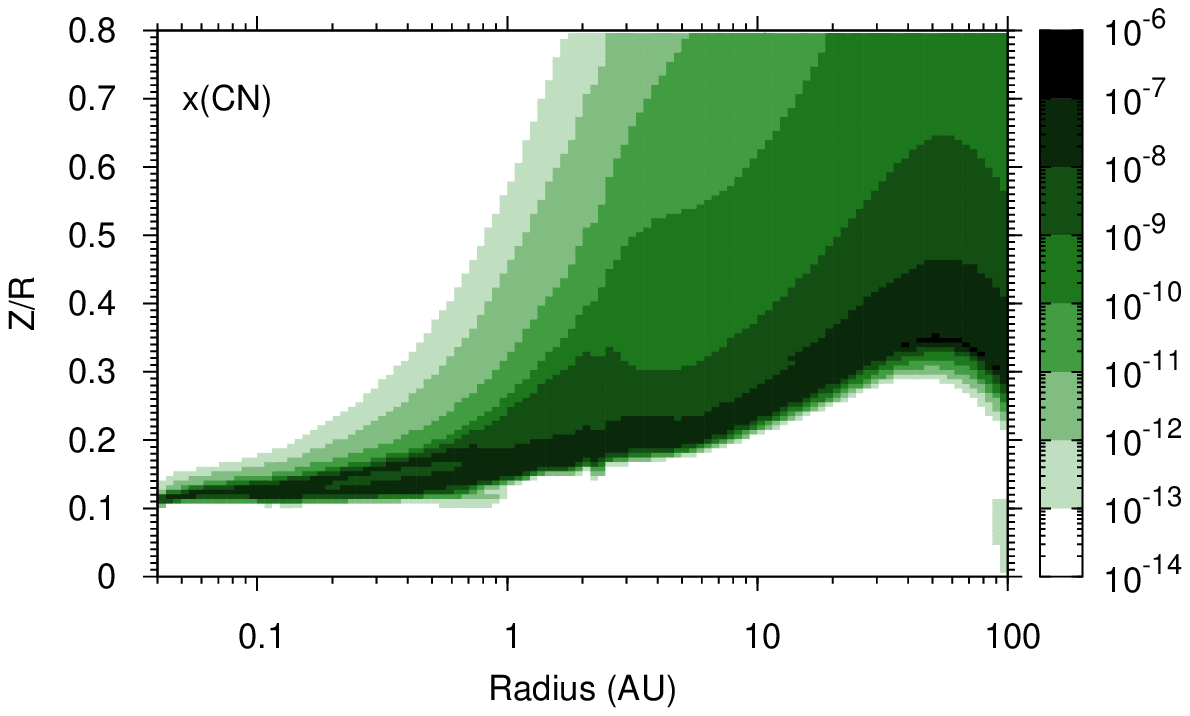}
\includegraphics[width=0.5\textwidth]{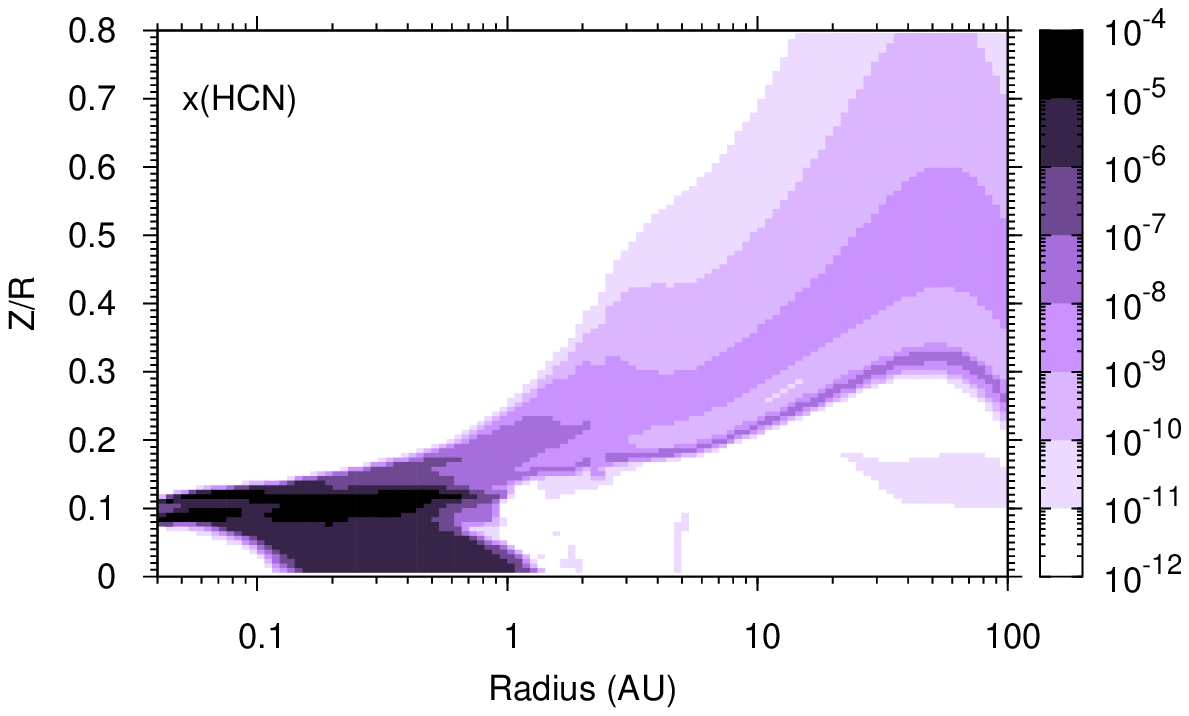}
\includegraphics[width=0.5\textwidth]{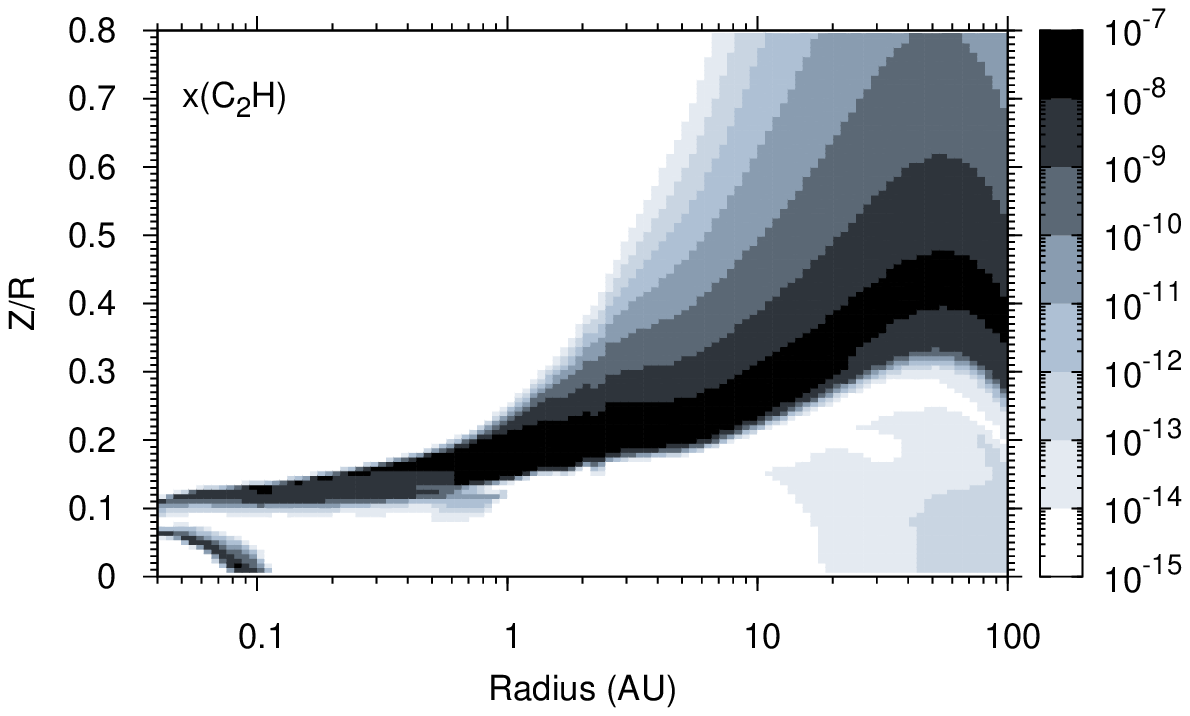}
\includegraphics[width=0.5\textwidth]{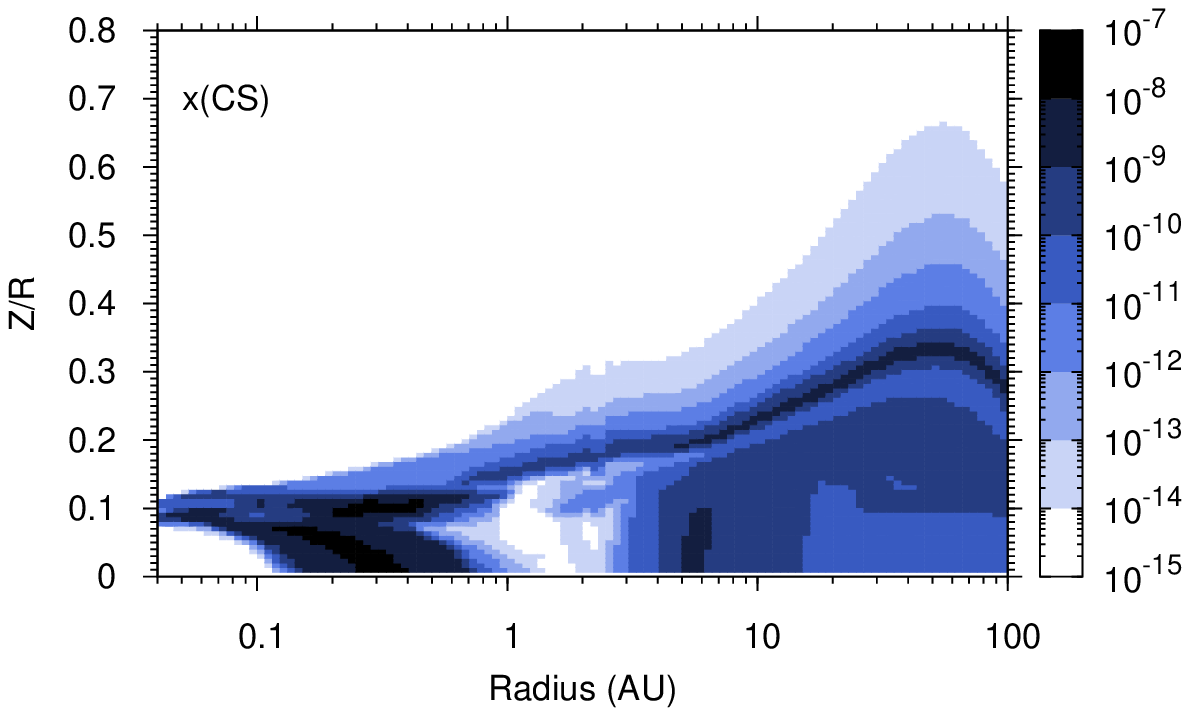}
\includegraphics[width=0.5\textwidth]{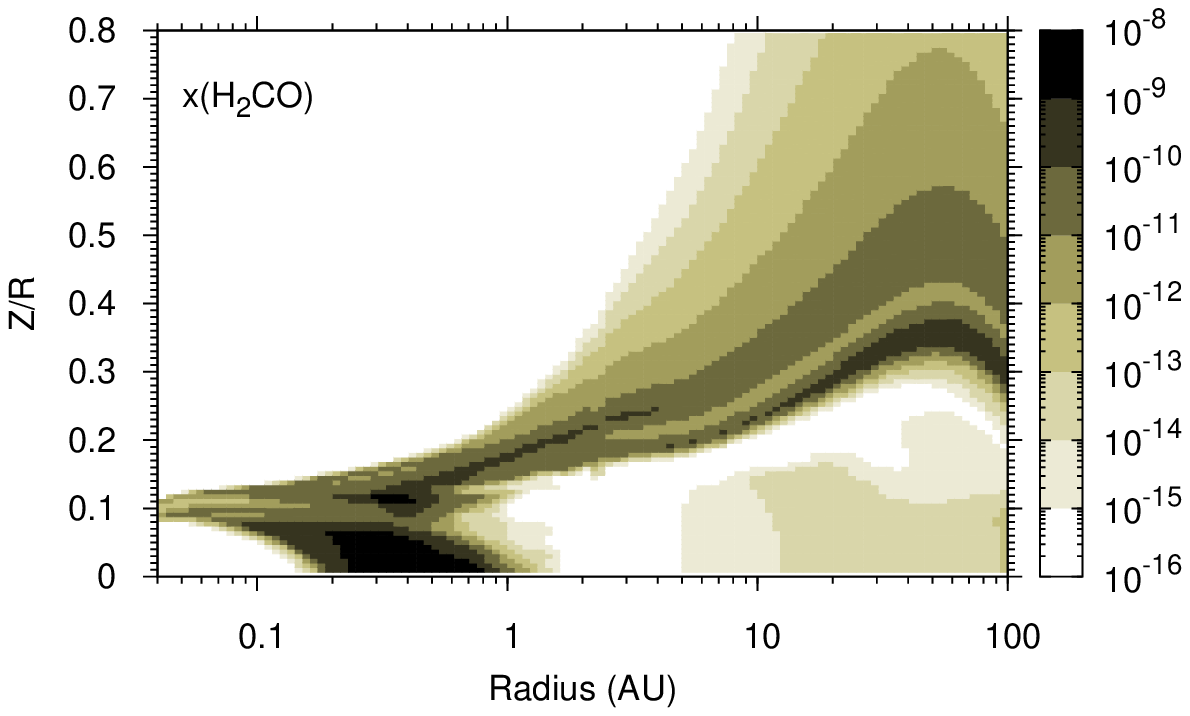}
\includegraphics[width=0.5\textwidth]{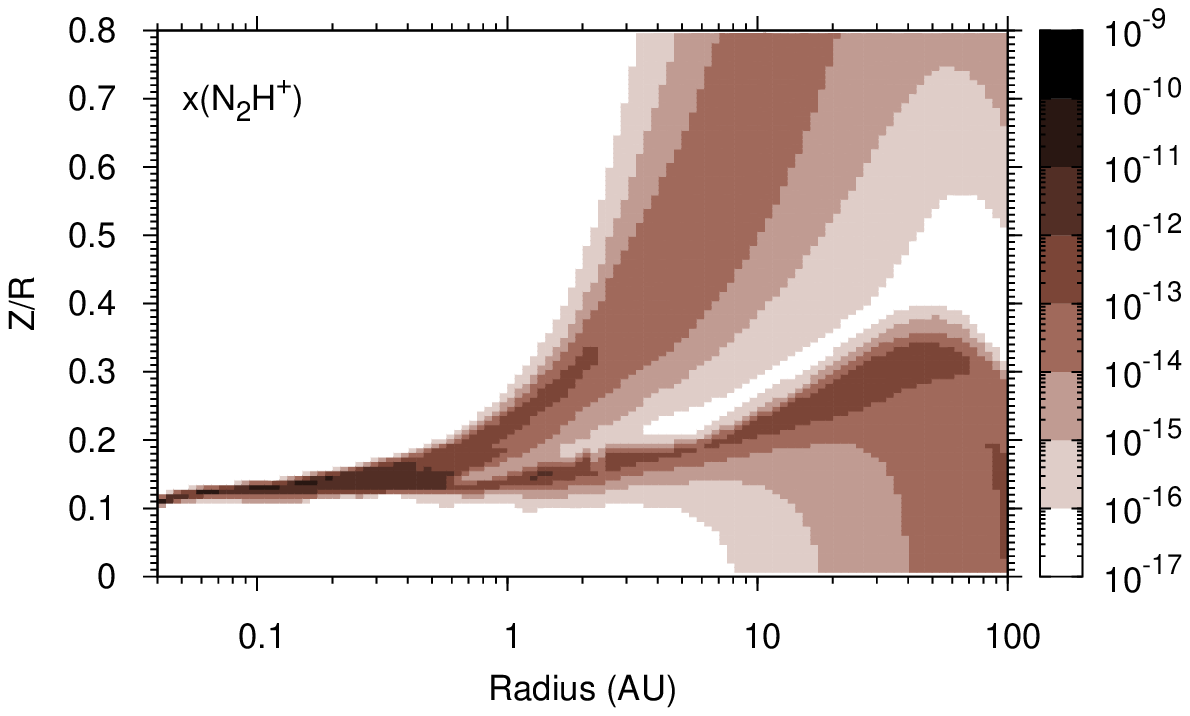}
\caption{Fractional abundance of molecules (relative to number density) 
as a function of disk radius, $R$, and height, $Z/R$.  }
\label{figure2}
\end{figure*}

\begin{figure}
\includegraphics[width=0.5\textwidth]{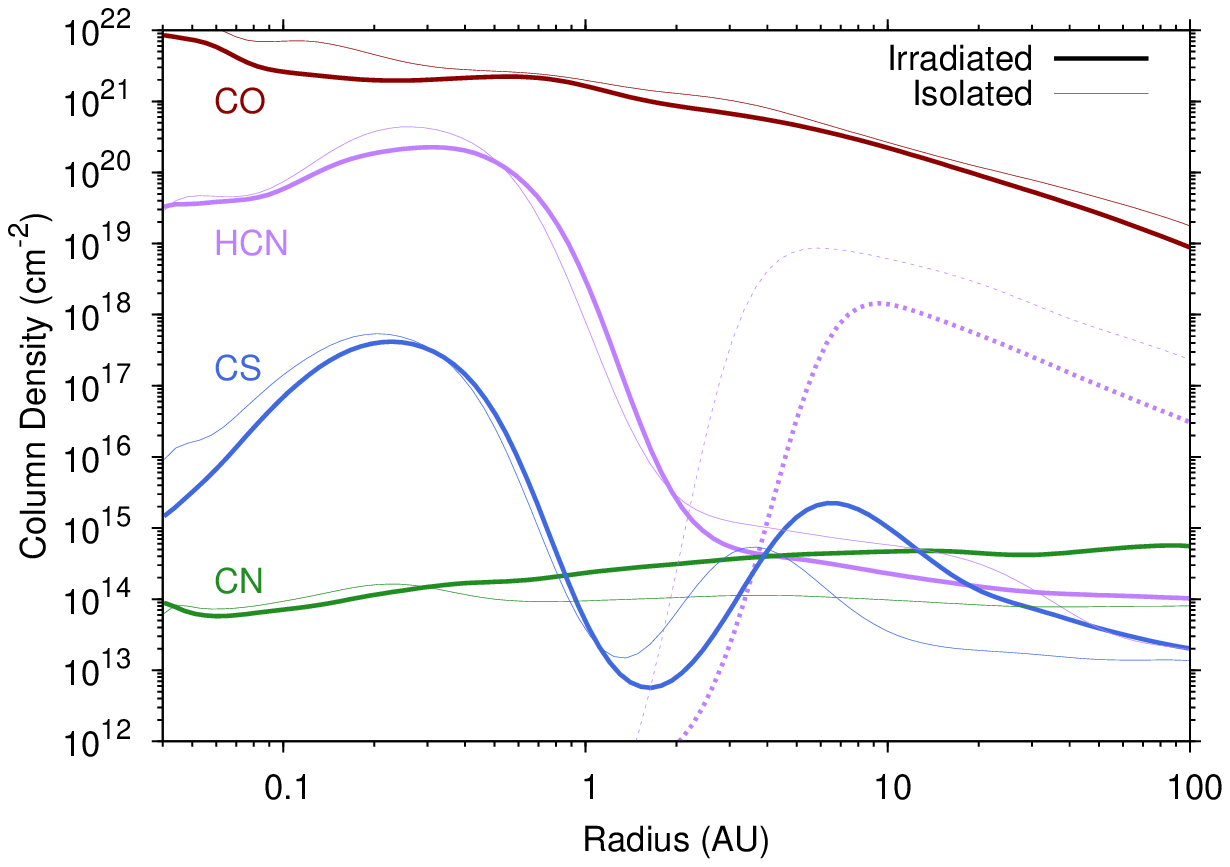}
\includegraphics[width=0.5\textwidth]{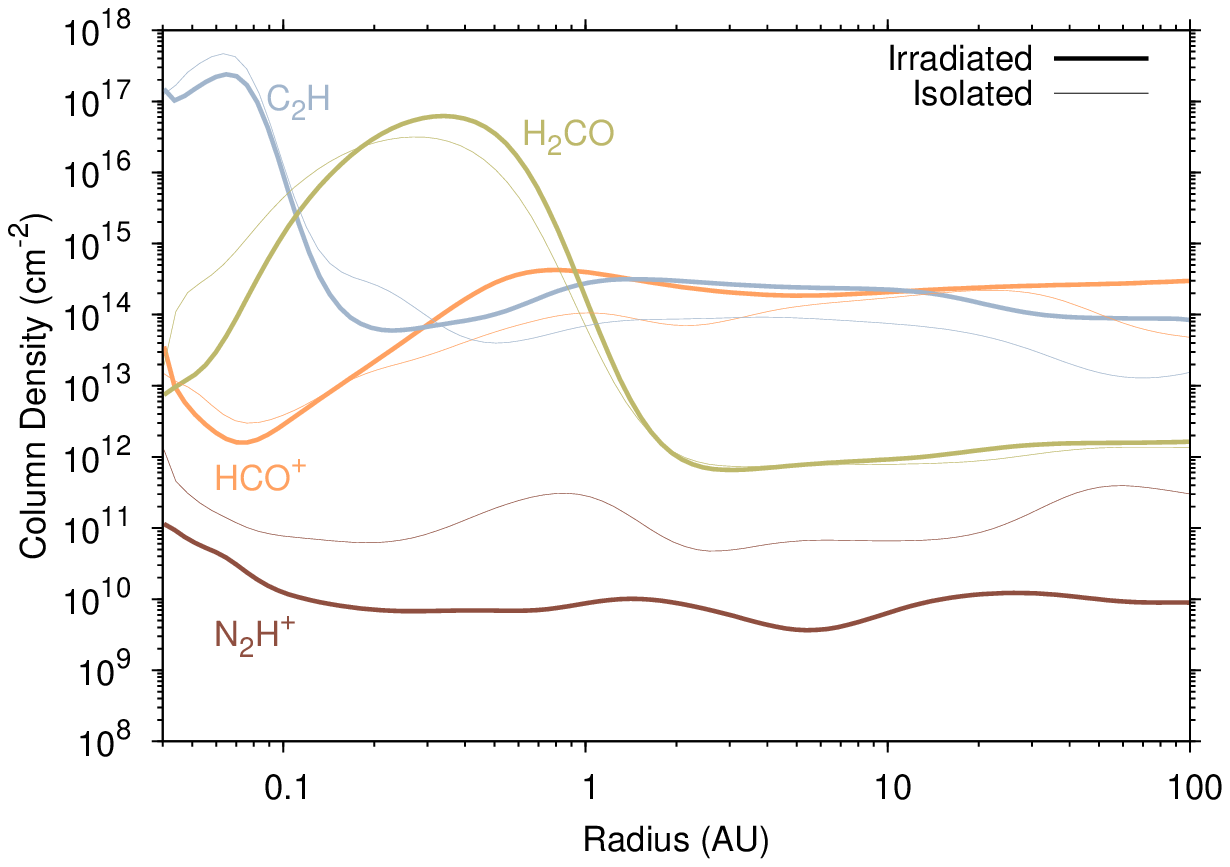}
\includegraphics[width=0.5\textwidth]{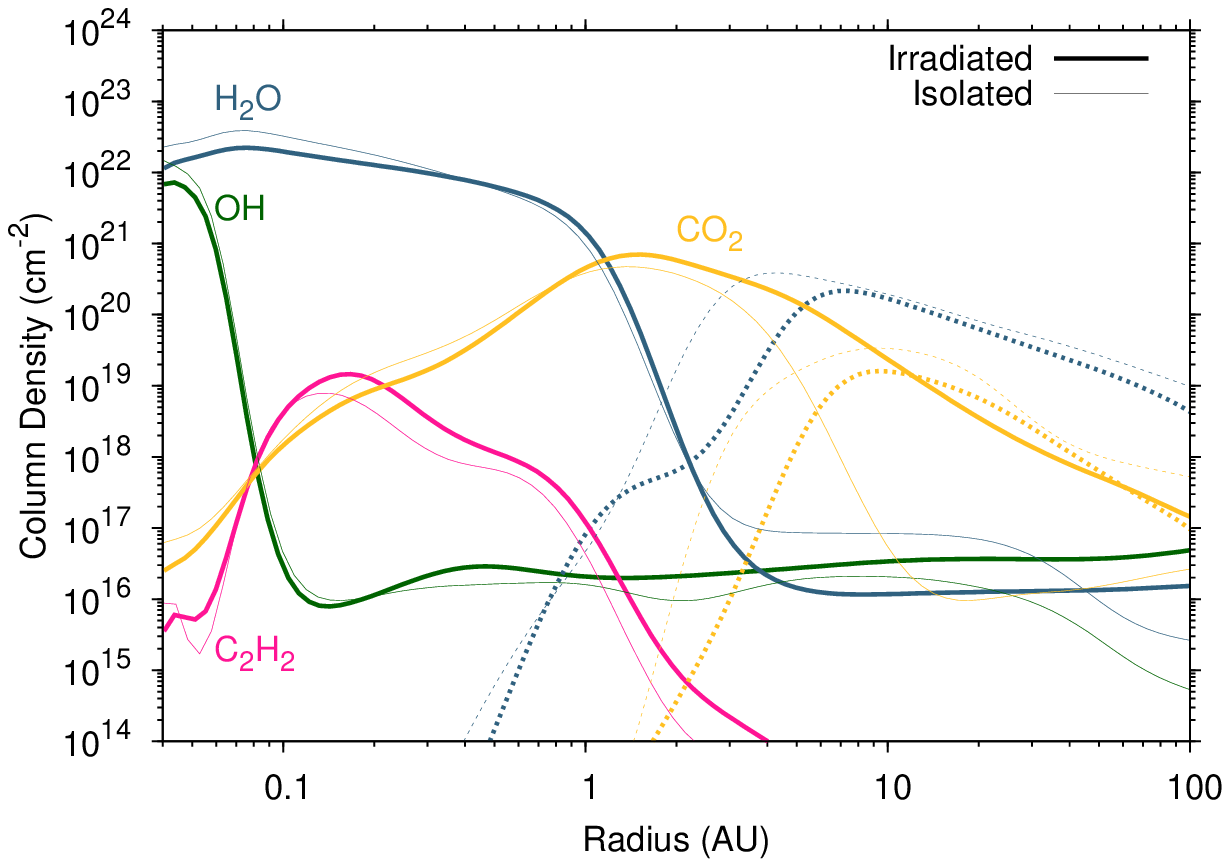}
\includegraphics[width=0.5\textwidth]{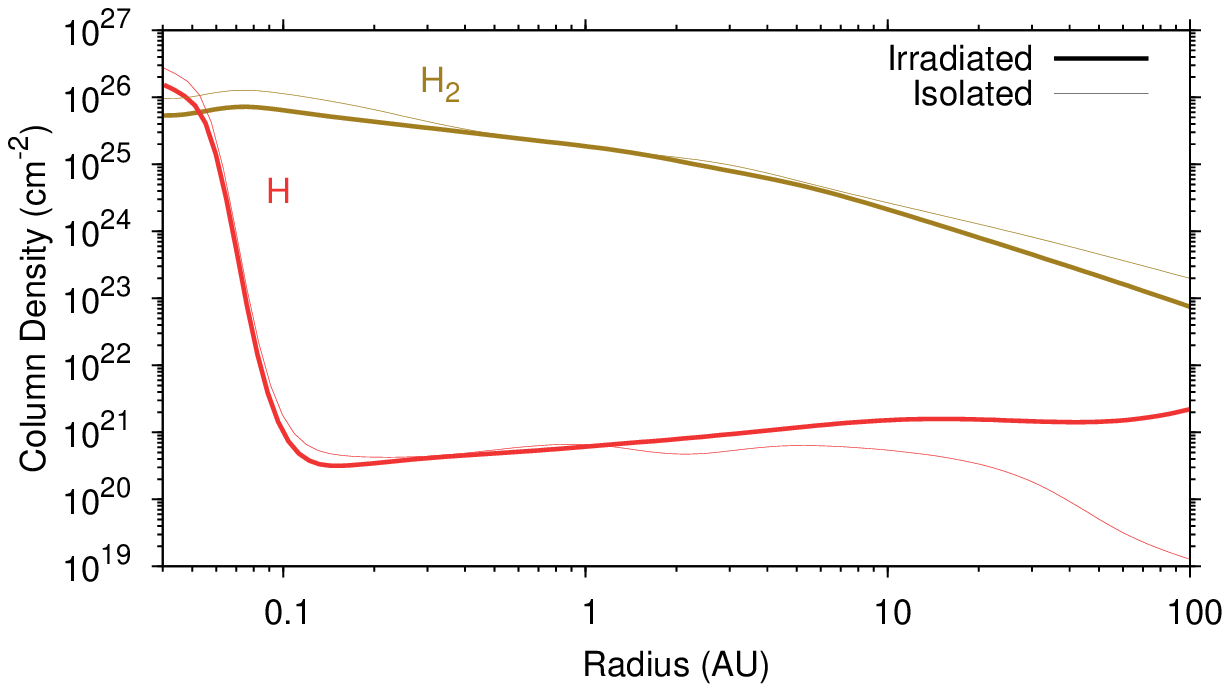}
\caption{Vertical column densities of species as a function of disk radius, $R$, 
for an irradiated disk (thick lines) and an isolated disk (thin lines).  
The grain-surface (ice) column densities are represented by the dashed lines. }  
\label{figure3}
\end{figure}

\begin{figure*}
\centering
\includegraphics[width=0.45\textwidth]{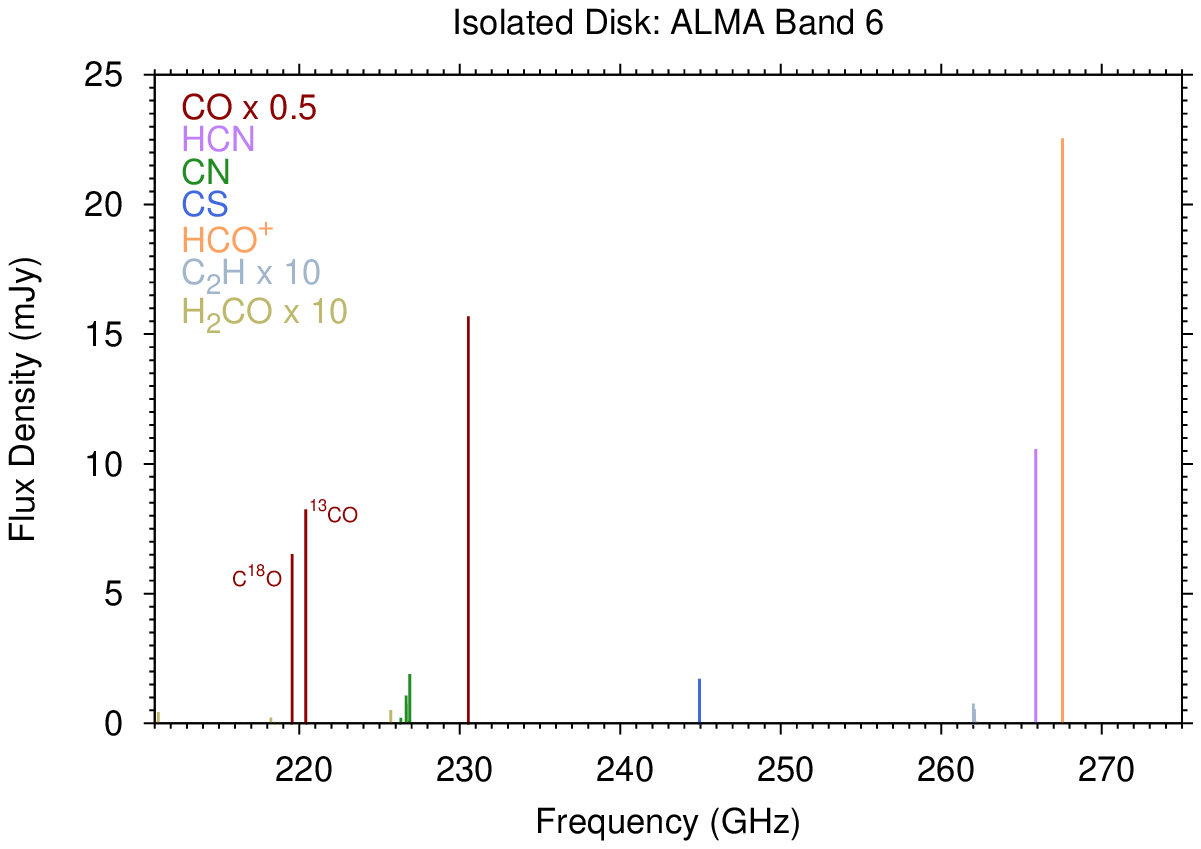}
\includegraphics[width=0.45\textwidth]{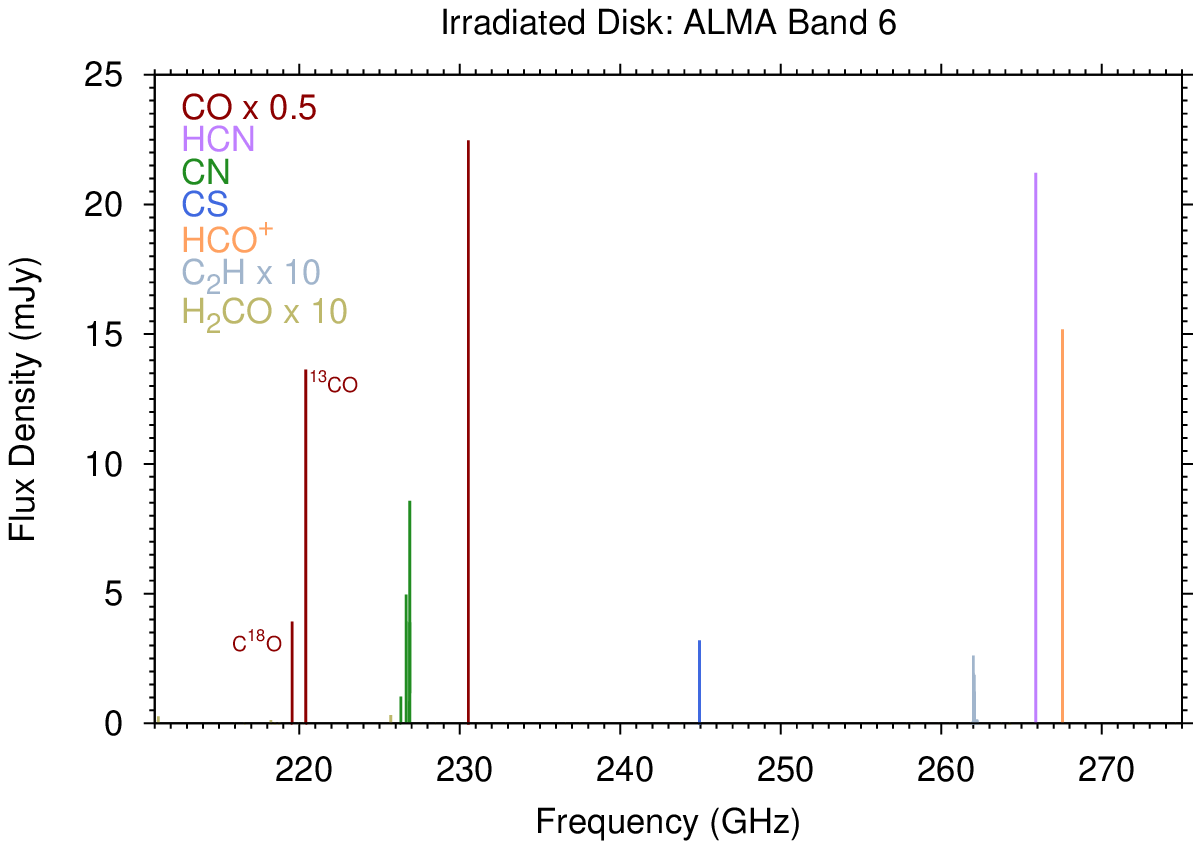}
\includegraphics[width=0.45\textwidth]{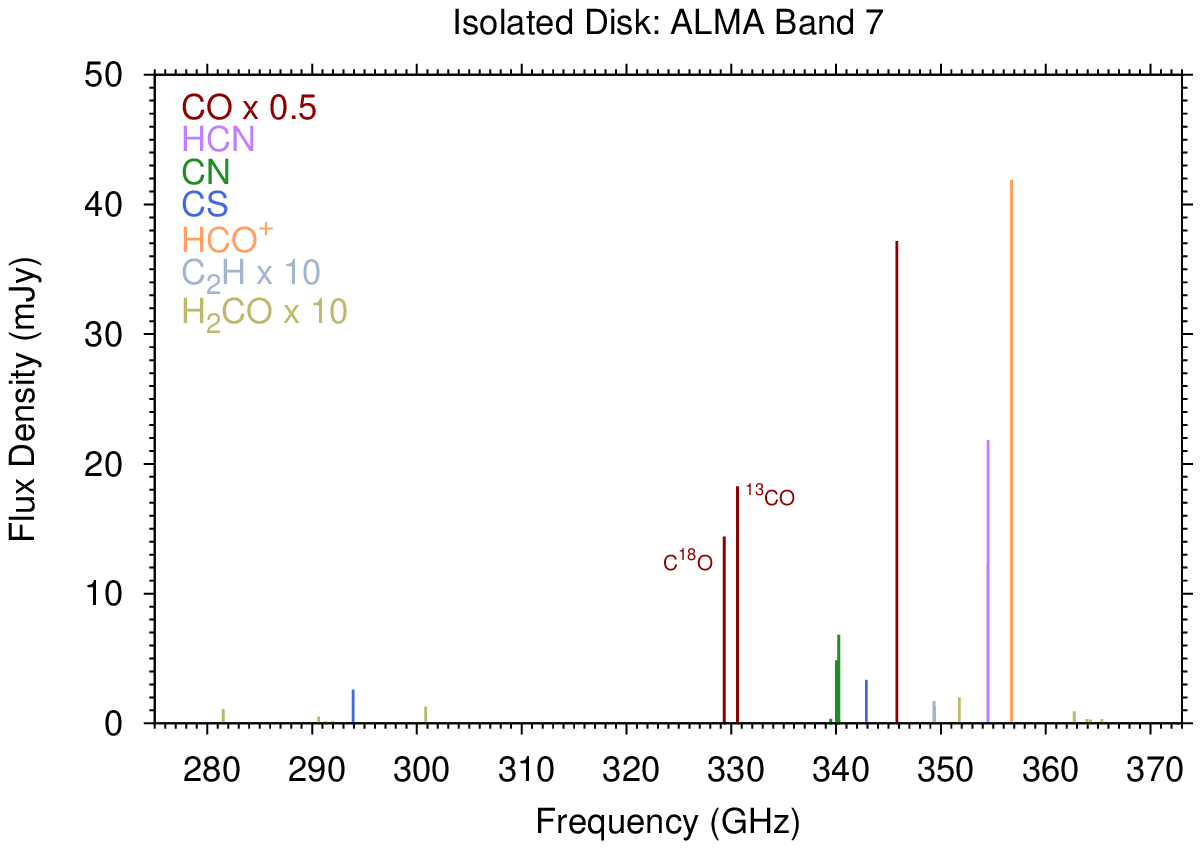}
\includegraphics[width=0.45\textwidth]{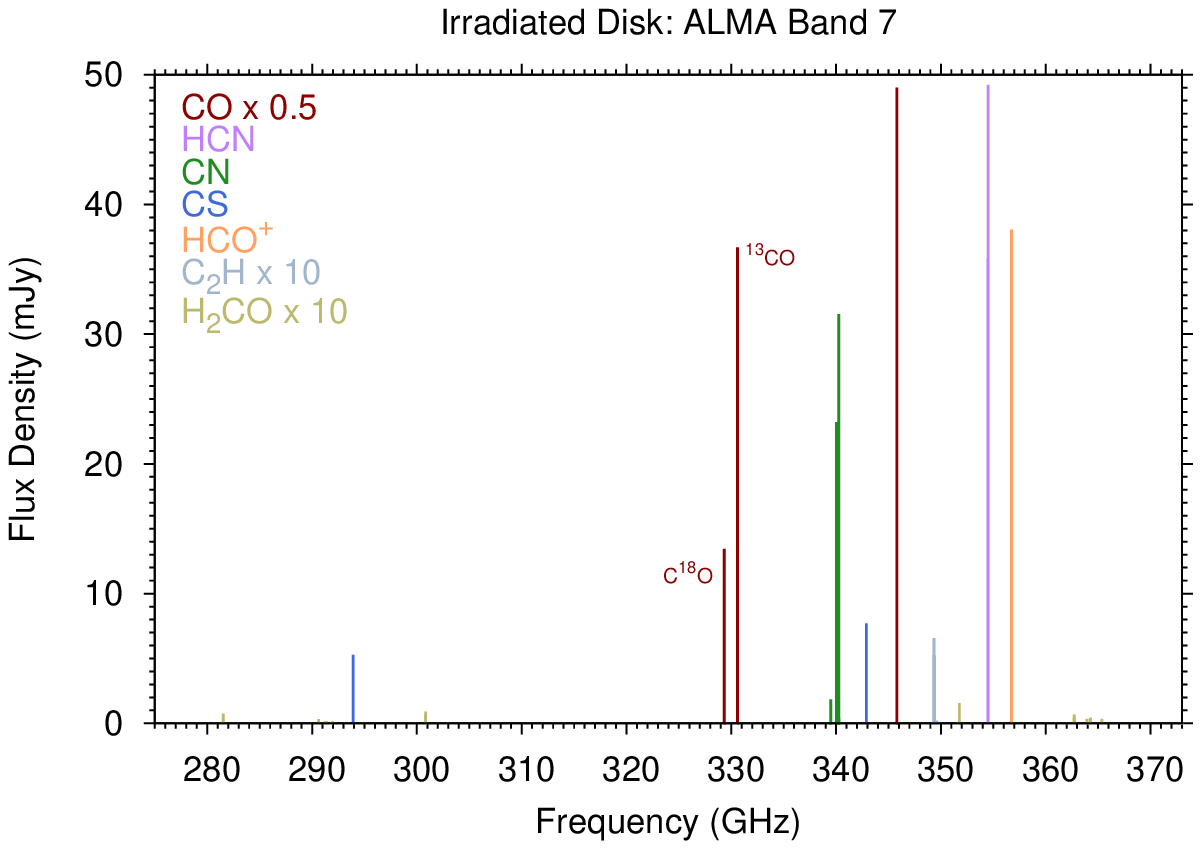}
\includegraphics[width=0.45\textwidth]{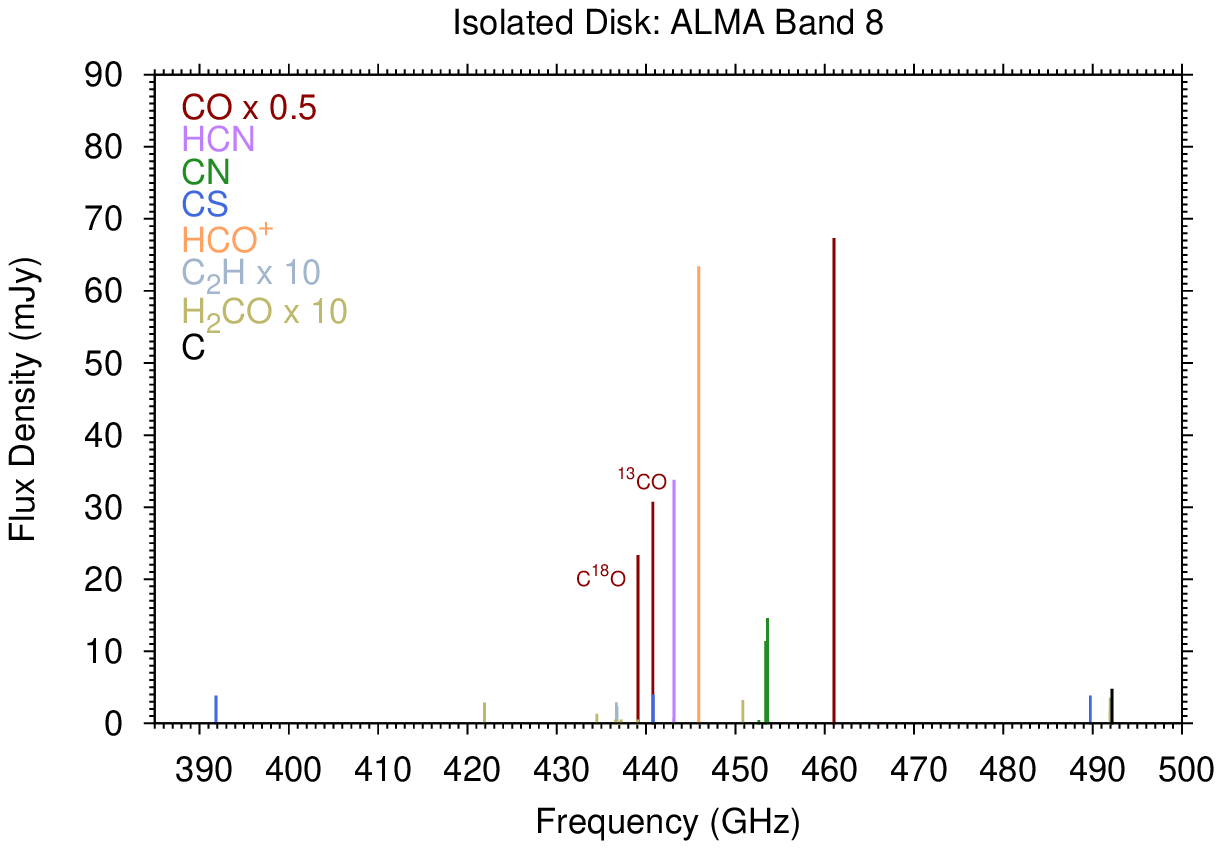}
\includegraphics[width=0.45\textwidth]{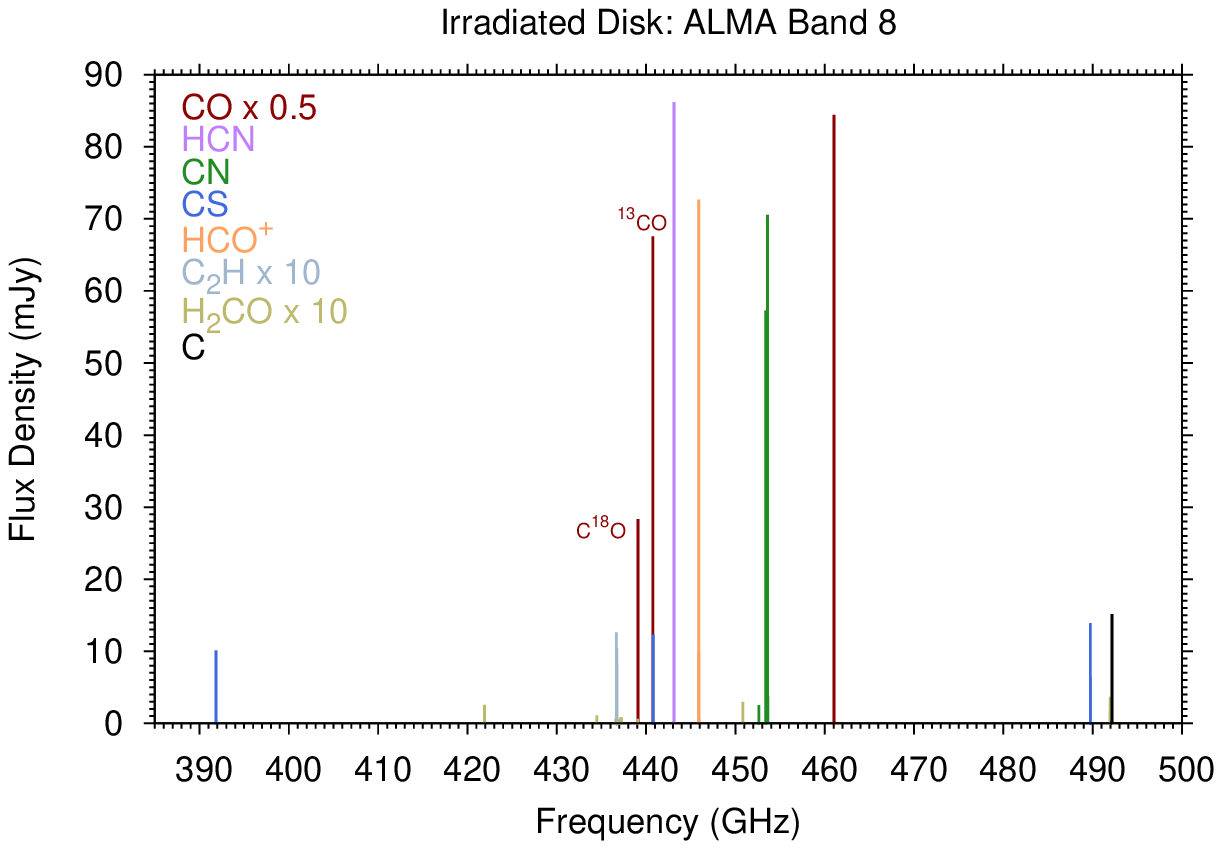}
\includegraphics[width=0.45\textwidth]{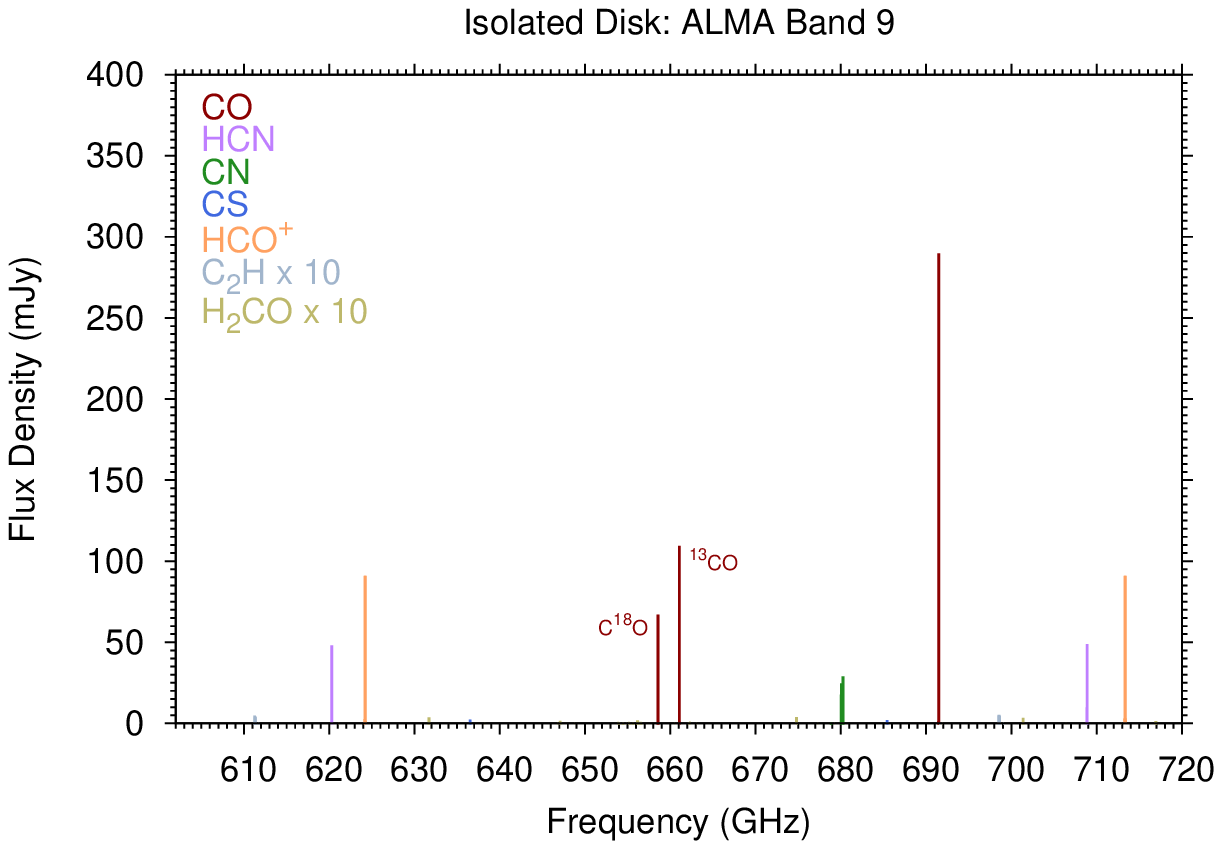}
\includegraphics[width=0.45\textwidth]{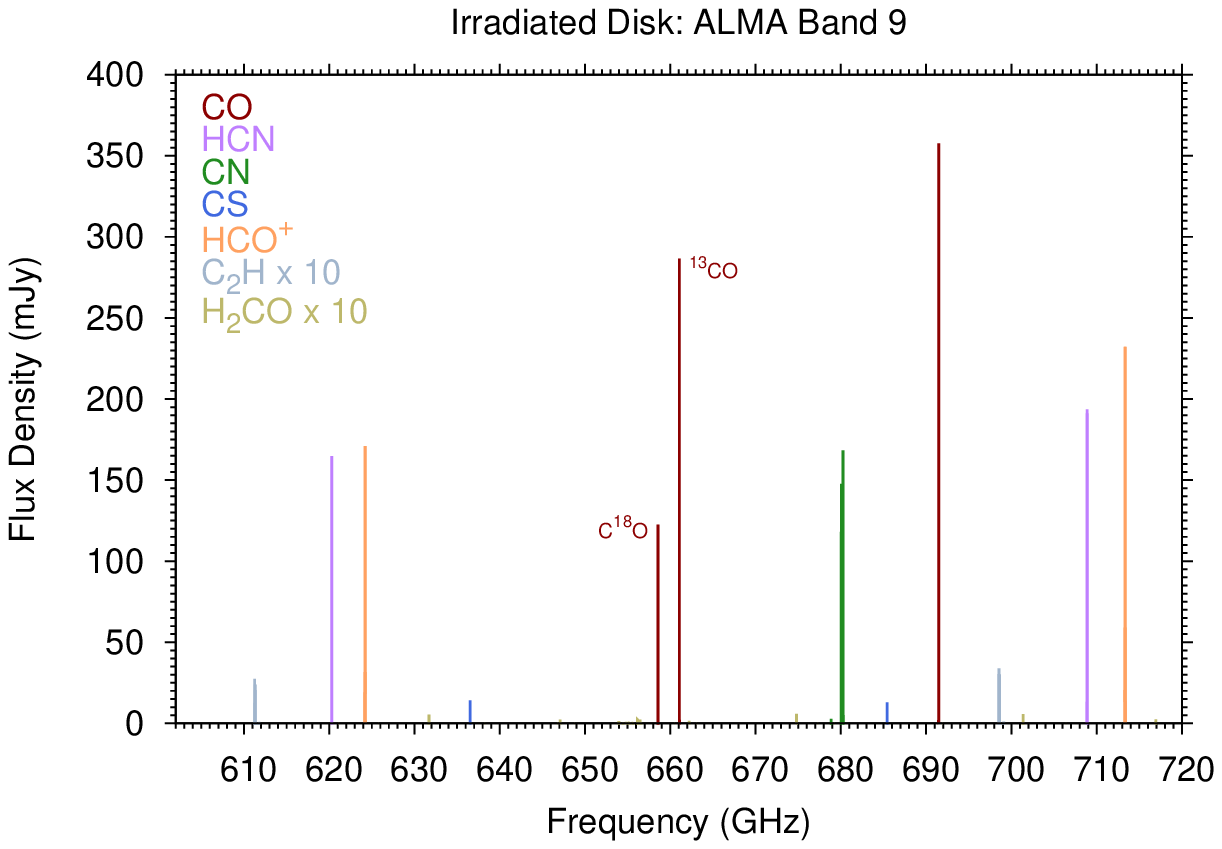}
\caption{Synthetic spectra in ALMA receiver bands 
6 through 9 for an isolated disk (left) and an irradiated disk (right). } 
\label{figure4}
\end{figure*}

Proplyds have been extensively studied since their 
discovery \citep[see, e.g.,][]{odell93,bally98},  
and were originally observed in silhouette against the bright nebular background. 
Early models concentrated on characterizing physical properties: 
morphology, surface density, mass, dust distribution, 
mass-loss rates, and lifetime 
\citep[see, e.g.,][]{johnstone98,storzer99,richling00,adams04,clarke07}.  
Models generally agree that the disk survives for a few Myr, beyond which, photoevaporation 
destroys the disk on a timescale $\sim$10$^5$~yr, depending on the distance from the massive star.   
In Figure~\ref{figure1}, we identify the region where the gas temperature equals the 
critical temperature for photoevaporation, i.e., where the photoevaporative flow likely originates.    
According to \citet{dullemond07}, $T_\mathrm{crit}$~$\approx$~0.2~$T_\mathrm{virial}$, where 
$T_\mathrm{virial}$ is the temperature at which the sound speed equals the escape velocity of the gas. 
Several groups \citep[e.g.,][]{adams04} model the disk as a photon-dominated region (PDR) including 
a small chemical network for the computation of the thermal balance and do not report 
the molecular structure beyond the transition from atomic to molecular hydrogen.  
Other authors report chemical structure calculations and resulting line emission; however, this 
has been restricted to ionic and atomic lines \citep[see, e.g.,][]{richling00}.  
\citet{nguyen02} studied the chemistry in irradiated disks around high-mass
and low-mass stars, modeling the disk structure as a series of one-dimensional vertical PDRs.  
They categorize ``observable" molecules as those 
possessing a value $\ge$10~$^{12}$~cm$^{-2}$.  
In their low-mass model, their column densities for 
HCN, H$_2$CO, H$_2$O, and C$_2$H are larger than ours by more than an 
order of magnitude despite both models generating similar CO column densities.  
We find the opposite case for HCO$^+$ and CN.  
These differences likely arise from different 
prescriptions for the disk structure: we perform a self-consistent calculation of the physical 
structure, whereas \citet{nguyen02} adopt assumed conditions.  
Our work highlights the importance of radiative transfer calculations in determining the potential 
observability of molecules, rather than relying on column density calculations.  

Here, we report the results of a self-consistent two-dimensional model 
of the physical and chemical structure of a protoplanetary disk irradiated by a nearby massive star.  
The disk remains predominantly molecular in nature over the lifetime of the 
disk ($\approx$10$^{6}$~yr) with potentially observable abundances of atoms and 
molecules (CI, CO, HCO$^+$, CN, and HCN) at (sub)mm wavelengths.  
The disk midplane is effectively shielded from UV radiation such that non-volatile 
molecules, e.g., HCN and H$_2$O, are frozen out onto dust grains  
suggesting icy planetesimal formation may be possible in proplyds.  
Observations of multiple transitions of these species (and isotopologues) 
would allow characterization of the gas surface density, excitation temperature, and optical depth.  
As far as we are aware, the work reported here is the first attempt to model the two-dimensional 
molecular structure and resultant (sub)mm molecular line emission of a proplyd.

\acknowledgments

Astrophysics at QUB is supported by a grant from the STFC.    
H.\ Nomura acknowledges the Grant-in-Aid for Scientific Research 21740137, 23103005 and the 
Global COE Program ``The Next Generation of Physics, Spun from Universality and 
Emergence'' from MEXT, Japan.

\clearpage


\clearpage


\begin{deluxetable}{lcccccccc}
\tablecaption{Column Densities and Integrated Line Strengths \label{table1}}
\tablewidth{0pt}
\tabletypesize{\small}
\tablehead{\colhead{Species} & \multicolumn{4}{c}{Column Density\tablenotemark{a}} &  \colhead{Transition} & \colhead{Frequency} 
& \multicolumn{2}{c}{Line Strength\tablenotemark{b}} \\
 & \multicolumn{4}{c}{(cm$^{-2}$)} & & (GHz) & \multicolumn{2}{c}{(mJy~km~s~$^{-1}$)} \\ 
 & \multicolumn{2}{c}{Isolated} & \multicolumn{2}{c}{Irradiated} & & & Isolated & Irradiated \\ 
 \cline{2-3}  \cline{4-5} \cline{8-9}\\
 & \colhead{10 AU} & \colhead{100AU} & \colhead{10 AU} & \colhead{100AU} & & & & \\ }
\startdata
H             & 5.6(20) & 1.3(19) & 1.6(21) & 2.2(21) & $\cdots$               & $\cdots$ & $\cdots$ & $\cdots$ \\
H$_2$         & 2.6(24) & 2.0(23) & 2.1(24) & 7.5(22) & $\cdots$               & $\cdots$ & $\cdots$ & $\cdots$ \\
C             & 2.3(17) & 7.2(16) & 4.7(17) & 2.3(17) & $^3$P$_1$--$^3$P$_0$   & 492.161  & $\cdots$ & 22.8     \\
              &         &         &         &         & $^3$P$_2$--$^3$P$_1$   & 809.342  & 24.8     & 85.2     \\
CO            & 2.7(20) & 1.8(19) & 2.3(20) & 8.8(18) & 1--0                   & 115.271  & 14.5     & 31.2     \\
              &         &         &         &         & 2--1                   & 230.538  & 68.2     & 141      \\
              &         &         &         &         & 3--2                   & 345.796  & 163      & 326      \\
              &         &         &         &         & 4--3                   & 461.041  & 293      & 578      \\
              &         &         &         &         & 6--5                   & 691.473  & 599      & 1230     \\
              &         &         &         &         & 7--6                   & 806.652  & 734      & 1600     \\
              &         &         &         &         & 8--7                   & 921.800  & 824      & 1970     \\
HCO$^+$       & 1.7(14) & 4.8(13) & 2.1(14) & 3.0(14) & 3--2                   & 267.558  & 27.1     & 31.6     \\
              &         &         &         &         & 4--3                   & 356.734  & 54.1     & 86.9     \\
              &         &         &         &         & 5--4                   & 445.903  & 86.4     & 184      \\
              &         &         &         &         & 7--6                   & 624.209  & 150      & 545      \\
              &         &         &         &         & 8--7                   & 713.342  & 183      & 830      \\
              &         &         &         &         & 9--8                   & 802.456  & 225      & 1200     \\
              &         &         &         &         &10--9                   & 891.558  & 284      & 1670     \\
CN            & 9.7(13) & 8.0(13) & 4.6(14) & 5.5(14) & 2$_{5/2}$--1$_{3/2}$   & 226.876  & $\cdots$ & 12.0     \\
              &         &         &         &         & 3$_{5/2}$--2$_{3/2}$   & 340.031  & $\cdots$ & 32.9     \\
              &         &         &         &         & 3$_{7/2}$--2$_{5/2}$   & 340.249  & $\cdots$ & 45.4     \\
              &         &         &         &         & 4$_{7/2}$--3$_{5/2}$   & 453.390  & 14.0     & 83.2     \\
              &         &         &         &         & 4$_{9/2}$--3$_{7/2}$   & 453.607  & 18.1     & 104      \\
              &         &         &         &         & 6$_{11/2}$--5$_{9/2}$  & 680.047  & 32.4     & 221      \\
              &         &         &         &         & 6$_{13/2}$--5$_{11/2}$ & 680.264  & 38.2     & 255      \\
              &         &         &         &         & 7$_{13/2}$--6$_{11/2}$ & 793.336  & 37.7     & 278      \\
              &         &         &         &         & 7$_{15/2}$--6$_{13/2}$ & 793.554  & 43.7     & 315      \\
              &         &         &         &         & 8$_{15/2}$--7$_{13/2}$ & 906.593  & 39.5     & 308      \\
              &         &         &         &         & 8$_{17/2}$--7$_{15/2}$ & 906.811  & 44.7     & 345      \\
HCN           & 5.9(14) & 2.1(13) & 2.4(14) & 1.0(14) & 3--2                   & 265.886  & 12.4     & 30.1     \\
              &         &         &         &         & 4--3                   & 354.505  & 27.1     & 71.9     \\
              &         &         &         &         & 5--4                   & 443.116  & 44.9     & 129      \\
              &         &         &         &         & 7--6                   & 620.304  & 80.2     & 251      \\
              &         &         &         &         & 8--7                   & 708.877  & 98.0     & 295      \\
              &         &         &         &         & 9--8                   & 797.433  & 120      & 316      \\
              &         &         &         &         &10--9                   & 885.971  & 149      & 313      \\
CS            & 2.7(13) & 1.5(13) & 1.2(15) & 2.0(13) & 7--6                   & 342.883  & $\cdots$ & 10.8     \\
              &         &         &         &         & 8--7                   & 391.847  & $\cdots$ & 14.2     \\
              &         &         &         &         & 9--8                   & 440.803  & $\cdots$ & 17.3     \\
              &         &         &         &         & 10--9                  & 489.751  & $\cdots$ & 19.6     \\
              &         &         &         &         & 13--12                 & 636.532  & $\cdots$ & 20.2     \\
              &         &         &         &         & 14--13                 & 685.436  & $\cdots$ & 18.5     \\
              &         &         &         &         & 17--16                 & 832.062  & $\cdots$ & 11.3     \\
C$_2$H        & 7.7(13) & 1.4(13) & 2.3(14) & 8.4(13) & $\cdots$               & $\cdots$ & $\cdots$ & $\cdots$ \\
H$_2$CO       & 8.1(11) & 1.4(12) & 9.3(11) & 1.6(12) & $\cdots$               & $\cdots$ & $\cdots$ & $\cdots$ \\
N$_2$H$^+$    & 6.7(10) & 3.1(11) & 7.0(09) & 8.9(09) & $\cdots$               & $\cdots$ & $\cdots$ & $\cdots$ \\
H$_2$O        & 8.4(16) & 2.6(15) & 1.2(16) & 1.5(16) & $\cdots$               & $\cdots$ & $\cdots$ & $\cdots$ \\
OH            & 2.2(16) & 5.2(14) & 3.5(16) & 4.9(16) & $\cdots$               & $\cdots$ & $\cdots$ & $\cdots$ \\
CO$_2$        & 1.3(16) & 2.7(16) & 2.4(19) & 1.4(17) & $\cdots$               & $\cdots$ & $\cdots$ & $\cdots$ \\
C$_2$H$_2$    & 2.7(14) & 3.0(13) & 2.8(13) & 1.2(13) & $\cdots$               & $\cdots$ & $\cdots$ & $\cdots$ 
\enddata
\tablenotetext{a}{~$a(b)$~=~$a$~$\times$~10$^{b}$.}
\tablenotetext{b}{~Listed line strengths are restricted to $>$10 mJy~km$^{-1}$.}
\end{deluxetable}

\end{document}